%% file: main.tex
\documentclass[%
reprint,
amsmath,
amssymb,
aps,
showkeys
]{revtex4-2}

\usepackage{graphicx}
\usepackage{dcolumn}
\usepackage{bm}
\usepackage[hidelinks]{hyperref}
\hypersetup{colorlinks,allcolors=blue}
\usepackage[braket,qm]{qcircuit}
\usepackage{float}
\usepackage{braket}
\usepackage{multirow}
\usepackage{mathrsfs}
\usepackage{soul}
\usepackage{booktabs}
\usepackage{xcolor}
\usepackage{tabularx}
\usepackage{enumitem}
\usepackage{tikz}
\usetikzlibrary{shapes,arrows,positioning,backgrounds}

\begin{document}

\title{Photonic Quantum Computers}

\author{Muhammad AbuGhanem{$^{1,2,\star}$}}
\address{$^{1}$ Faculty of Science, Ain Shams University, Cairo, 11566, Egypt}
\address{$^{2}$ Zewail City of Science, Technology and Innovation, Giza, 12678, Egypt}

\email{gaa1nem@gmail.com}

\date{\today}

\begin{abstract}

In the pursuit of scalable and fault-tolerant quantum computing architectures, photonic-based quantum computers have emerged as a leading frontier. This article provides a comprehensive overview of advancements in photonic quantum computing, developed by leading industry players, examining current performance, architectural designs, and strategies for developing large-scale, fault-tolerant  photonic quantum computers. It also highlights recent groundbreaking experiments that leverage the unique advantages of photonic technologies, underscoring their transformative potential. This review captures a pivotal moment of photonic quantum computing in the noisy intermediate-scale quantum (NISQ) era, offering insights into how photonic quantum computers might reshape the future of quantum computing.

\end{abstract}


\maketitle

\tableofcontents

\section{Introduction}

In the ever-evolving landscape of technology, quantum computers stand out as revolutionary and highly promising entities~\citep{NISQ23,DiVincenzo,qc,Nilson}, extending their influence from cryptography and artificial intelligence to drug discovery, optimization problems and communications~\citep{NISQ23}. By harnessing the fundamental principles of quantum mechanics~\citep{principles,QCModels}, quantum computers possess the capacity to address intricate problems with unparalleled speed, surpassing the computational capabilities inherent in the most powerful classical computer known at present~\citep{NISQ23,nisqQC10,art19,Google_superm_2023,GSA,Jiuzhang,Jiuzhang2.0,Jiuzhang3.0,Borealis5,Zuchongzi,Zuchongzi2.1,Googles,Sycamore}.

Photonic quantum computers are currently prominent contenders in fault-tolerant quantum computation (FTQC). These advanced architectures utilize photons as the medium for qubit encoding and manipulation~\citep{PsiQ8}, exhibiting inherent resilience against decoherence and noise, even at room temperature. This makes them exceptionally well-suited for scalable and FTQC. Photonic quantum computing also stands out for enabling the construction of modular, easily networked quantum computers, holding significant potential for practical applications~\citep{NISQ23}.

A key characteristic of photonic quantum computing lies in the encoding of qubits within the quantum state of light, unlocking a multitude of possibilities for quantum information processing (QIP). Quantum states of light have played a pivotal role since the inception of groundbreaking experiments in nonlocality and quantum teleportation~\citep{IntegPhot1,IntegPhot2}. The benefits resulting from the encoding of qubits in single-photon states are numerous, including immunity to decoherence phenomena, the feasibility of conducting information processing at room temperature, and the capability to transmit photons through both optical fibers and free-space channels.

Effective photonic QIP requires photonic processors to meet four key requirements. \textit{Firstly,} they must be large-scale to handle complex problem-solving. \textit{Second,} universality is essential, enabling the implementation of arbitrary transformations that map the system onto various problems. Achieving universality requires all-to-all connectivity ($n$ inputs to $n$ outputs) and full reconfigurability~\citep{Univ12-mod1,Univ12-mod25,Univ12-mod26,Univ12-mod37}. \textit{Third,} low loss is crucial to preserve the integrity of (quantum) information carried by the system. \textit{Finally,} it is essential for a photonic processor to maintain quantum interference, ensuring the accuracy and reliability of quantum computations~\citep{Univ12-mod}.

The integration of photonics into quantum computing holds transformative potential across various fields. 
In the last decade, significant strides in photonic quantum technologies have led to heightened system complexity, resulting in breakthroughs across diverse facets of quantum information science. Noteworthy achievements include the realization of quantum advantage~\citep{Jiuzhang,Jiuzhang2.0,Borealis} and the establishment of satellite quantum communications~\citep{IntegPhot6,IntegPhot7,BSBN4600}. 
More recently, photonic processors have garnered increasing attention due to their versatile applications. 
These applications span QIP based on linear optics~\citep{Univ12-mod1,Univ12-mod2,Univ12-mod3,Univ12-mod4,Univ12-mod5,Univ12-mod6,Univ12-mod7,Univ12-mod8,Univ12-mod9,Univ12-mod10,Univ12-mod11,Univ12-mod12,Univ12-mod13,Univ12-mod27,Univ12-mod28,PhoInformProc224}, quantum machine learning (QML)~\citep{Univ12-mod18,Univ12-mod19,Univ12-mod20,Univ12-mod21,Univ12-mod22}, quantum repeater networks~\citep{Univ12-mod14,Univ12-mod15,Univ12-mod16,Univ12-mod17}, and radio-frequency signal processing~\citep{Univ12-mod23,Ipronics1}.

Functioning as tunable multimode interferometers capable of executing arbitrary linear optical transformations, photonic processors have been realized in various topologies, including triangular~\citep{Univ12-mod1,Univ12-mod25}, square~\citep{Univ12-mod26}, hexagonal~\citep{Ipronics1}, fan-like~\citep{Univ12-mod19}, rhomboidal~\citep{Univ12-mod4}, and quadratic configurations~\citep{Univ12-mod23}.
Photonic quantum computers are set to significantly enhance computational efficiency, surpassing classical systems in areas such as cryptography~\citep{QCinternet133}, advancing quantum chemistry and materials science~\citep{Univ12-mod8,XanChem1,XanChem2}. These machines have the potential to revolutionize secure communication~\citep{QCinternet154}, refine molecular simulations critical for drug discovery~\citep{XanChem5,XanChem8}, and optimize complex logistical networks~\citep{QCinternet201}.

In this paper, we provide an in-depth review of key players in photonic quantum computing technologies (listed alphabetically) and their developed quantum photonic computers, delineating their efforts towards the realization of fault-tolerant photonic quantum computers. 
Featuring noteworthy contributions from key players such as 
iPronics, USTC Jiuzhang, ORCA Computing, Photonic Inc., PsiQuantum, Quandela Photonic Quantum Computers, QuixQuantum, TundraSystems, 
TuringQ, and  Xanadu, among others. We also explore the performance of these photonic quantum processors, our focus extends to recent experiments that leverage the quantum computational advantages inherent in photonic-based quantum computers, highlighting the substantial contributions of these entities to the field. 
For a detailed exploration of photonic quantum computing, including critical aspects such as photon encoding, device components, photonic quantum communication and internet, and programmable photonic integrated circuits (PPICs), we refer to~\citep{Photonics1}. This review also covers applications, prospects, and challenges in photonic quantum computing, which extend beyond the scope of our current review.

\section{iPronics Programmable Photonics}

\subsection{Overview}

Established in 2019, iPronics~\citep{Ipronics} is dedicated to pushing the boundaries of versatile integrated programmable photonic systems~\citep{Ipronics3,Multipurpose1,Multipurpose2,Multipurpose3,Multipurpose4,Multipurpose5,Multipurpose6,Multipurpose7,Multipurpose8,Multipurpose9,Multipurpose10,Univ12-mod23,Multipurpose40}, wherein the synergy of optical hardware and software facilitates multifaceted functionalities. These systems find applications across diverse domains, including RF signal processing, optical communications~\citep{Multipurpose11,Multipurpose12,Multipurpose13,Multipurpose14,Multipurpose15}, sensing and biophotonics~\citep{Multipurpose18,Multipurpose19}, artificial intelligence, machine learning techniques~\citep{self21}, hardware acceleration~\citep{Multipurpose31}, neuromorphic computing, and quantum computing~\citep{NISQ23,IpronicsBOOK}.

\subsection{iPronics \textit{SmartLight} processor}

The iPronics \textit{SmartLight} processor is characterized by its silicon photonics chip operating in the C-band, featuring a hexagonal mesh configuration comprising 72 tuning units and 64 input/output ports~\citep{iPronicsProcessor}. Notable performance attributes encompass high-capacity building blocks designed for 50, 100, and 200 GHz bandwidth filtering functions, affording programmable control over central wavelength and extinction ratio, as well as demultiplexing capabilities. The processor is equipped with a laser spanning a 2nm range, with a central wavelength at 1550 nm. Augmenting user accessibility, the associated software exhibits a user-friendly interface with automated functions, while an integrated driving unit ensures comprehensive operational control~\citep{iPronicsProcessor,Univ12-mod5,Multipurpose33,Multipurpose34,Multipurpose35,PPCs8,Multipurpose37,Univ12-mod26}.

\subsection{Reconfigurable photonic processors}

\begin{figure*}
    \centering
    \includegraphics[width=0.9\textwidth]{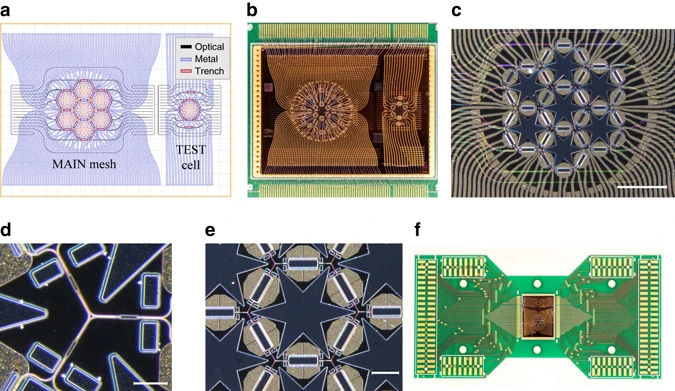}
    \caption{
    The hexagonal waveguide mesh chip was manufactured with the following key features: (a) Design layers, encompassing optical, electrical, and thermal components, were incorporated for both the 7-cell hexagonal waveguide mesh and an auxiliary test cell. (b) The silicon on insulator (SOI) chip, with dimensions of $15\times 20$ mm, was fabricated. (c) A detailed view of the 7-cell hexagonal waveguide mesh, with a scale bar of 2 mm. (d) An enlarged image of an optical interconnection node featuring three tunable basic units (TBUs), with a scale bar of 100 $\mu$m. (e) A close-up image of a single hexagonal cell exhibiting the Mach-Zehnder Interferometer (MZI), with a scale bar of 500 $\mu$m, and additional elements such as tuning heaters and star-type thermal isolation trenches in the right bottom corner. (f) The printed circuit board, where the waveguide mesh chip is mounted and wire bonded for integration. Reproduce from Ref.~\citep{Ipronics1} under a Creative Commons Attribution 4.0 International License (\url{ http://creativecommons.org/licenses/by/4.0/}).}
    \label{fig:iProniChip}
\end{figure*}

The research conducted by iPronics marks toward progress in advancing the realization of an innovative paradigm characterized by versatile and reconfigurable photonic processors~\citep{Multipurpose1,Multipurpose2,Multipurpose3,Multipurpose4,Multipurpose5,Multipurpose6,Multipurpose7,Multipurpose8,Multipurpose9,Multipurpose10}. 
In~\citep{Ipronics1}, Pérez \textit{et al.} has designed, fabricated, and exhibited an integrated reconfigurable photonic signal processor~\citep{Multipurpose1,Multipurpose2,Multipurpose3,Multipurpose4,Multipurpose5,Multipurpose6,Multipurpose7,Multipurpose8,Multipurpose9,Multipurpose10}. The chip employs a configurable hexagonal silicon waveguide mesh structure, with each hexagonal side comprising two waveguides~\citep{Multipurpose41} that can be dynamically coupled or switched using a programmable tuneable basic unit (TBU) implemented through a Mach-Zehnder interferometer (MZI)~\citep{PPCs8,Multipurpose37,Univ12-mod26,Multipurpose41,Multipurpose42,Multipurpose43}. iPronics systematically modified a 7-hexagon cell design to empirically manifest $21$ distinct functionalities, encompassing asymmetrical FIR Mach-Zehnder filters~\citep{Multipurpose42,Multipurpose43}, ring cavities, complex CROW~\citep{Multipurpose47}, SCISSOR~\citep{Multipurpose48}, and ring-loaded MZI filters~\citep{Multipurpose42,Multipurpose44,Multipurpose49}, as well as multiple input multiple output linear optic transformation devices (see Figure~\ref{fig:iProniChip}), notably a CNOT gate~\citep{Ipronics1}. These devices find application across diverse domains such as communications~\citep{Multipurpose11,Multipurpose12,Multipurpose13,Multipurpose14,Multipurpose15}, biophotonics and biomedical sensing~\citep{Multipurpose18,Multipurpose19},  multiprocessor and memory units interconnection~\citep{Multipurpose25,Multipurpose26,Multipurpose27}, high-speed signal processing operations~\citep{Multipurpose20,Multipurpose21,Multipurpose22,Multipurpose23,Multipurpose24}, advanced civil radar systems~\citep{Multipurpose17} internet of things, 
quantum logic gates~\citep{Multipurpose28,Multipurpose29,Multipurpose30}, and quantum information~\citep{Ipronics1}.

\begin{figure*}
    \centering
    \includegraphics[width=\textwidth]{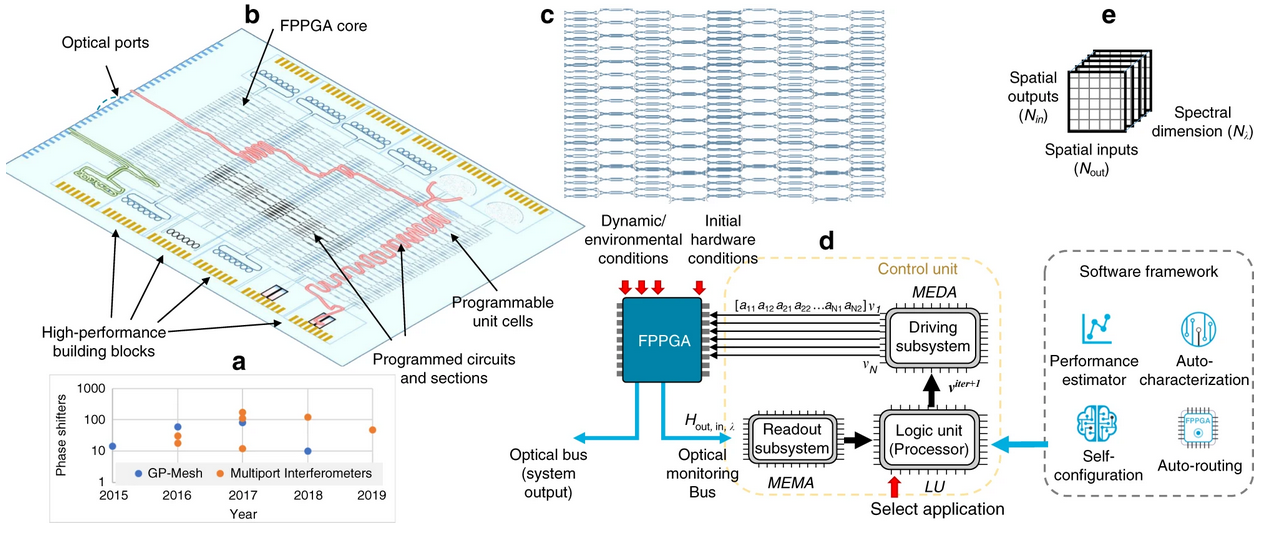}
    \caption{\textbf{Integrated Photonic hardware and control architecture of multiuse programmable photonic circuits.}
(a) The quantity of integrated phase shifters observed in recent waveguide mesh circuits. (b) The architecture of a labeled field programmable photonic gate array (FPPGA), comprising a waveguide mesh core and high-performance building blocks. (c) The FPPGA core incorporates a longitudinally parallel hexagonal waveguide mesh interconnection topology~\citep{Ipronics2}. (d) The electronic control subsystem, signals, and software procedures are depicted, illustrating the mechanisms for controlling the programmable photonic integrated circuit. (e) The data array presents the complete scattering matrix of the FPPGA core, encompassing input and output spatial ports, along with the optical spectral dimension. Key components include GP (general-purpose), MEMA (multichannel electrical monitoring array), MEDA (multichannel electrical driving array), and LU (logic unit). 
Reproduced from Ref.~\citep{Ipronics2} under a Creative Commons License (\url{http://creativecommons.org/licenses/by/4.0/}).}
    \label{fig:iProniInteG}
\end{figure*}

Additionally, Pérez-López \textit{et al.} introduced a control architecture and a series of control strategies~\citep{self19,self20,self21} designed for fault-tolerant self-configuration~\citep{self23} of the circuit to execute specific tasks in~\citep{Ipronics2}, as shown in Figure~\ref{fig:iProniInteG}. These algorithms are categorized into configuration methods that necessitate pre-characterization routines~\citep{self28}, as also outlined, and advanced optimization methods that not only eliminate this requirement but also address challenges arising from nonideal components characterized by nonhomogeneous loss distribution, power consumption~\citep{self26,self27}, phase offsets, optical crosstalk, and tuning crosstalk~\citep{self32,self33,self34,self35,Multipurpose22}. The authors further proposed and implemented self-configuration routines based on stochastic population-based methods, exemplified by Genetic Algorithms (GAs)~\citep{self20} and Particle Swarm Optimization (PSO)~\citep{self49}, for three distinct applications: an all-cross router, a beamsplitter, and an optical filter adaptable to a diverse range of spectral masks~\citep{Ipronics2}. Through the amalgamation of computational optimization and photonics, this endeavor marks a noteworthy stride toward a novel paradigm in programmable photonics~\citep{Ipronics2,IpronicsBOOK}.

The advancing maturity of integrated photonic technology has enabled the construction of progressively larger and intricate photonic circuits directly on chip surfaces~\citep{Multipurpose1,Multipurpose2,Multipurpose3,Multipurpose4,Multipurpose5,Multipurpose6,Multipurpose7,Multipurpose8,Multipurpose9,Multipurpose10}. While the majority of these circuits are presently tailored for specific applications, the heightened complexity has ushered in a new era of programmable photonic circuits~\citep{IpronicsBOOK,self8,self9,Ipronics3,Ipronics5}. These circuits can be dynamically configured using software, leveraging a network of on-chip waveguides, tunable beam couplers~\citep{self12,Univ12-mod37,self40}, and optical phase shifters~\citep{self37,self38} to serve a diverse array of functions~\citep{Multipurpose44}. Further comprehensive insights into the research potential of iPronics in this domain can be explored in~\citep{IpronicsBOOK,Ipronics1,Ipronics2,Ipronics3,Ipronics4,Ipronics5,IpronicWhite}.

\section{The \textit{Jiuzhang} Photonic Quantum Computers}\label{Jiuzhang}

\begin{figure*}
    \centering
   \includegraphics[width=\textwidth,height=12cm]{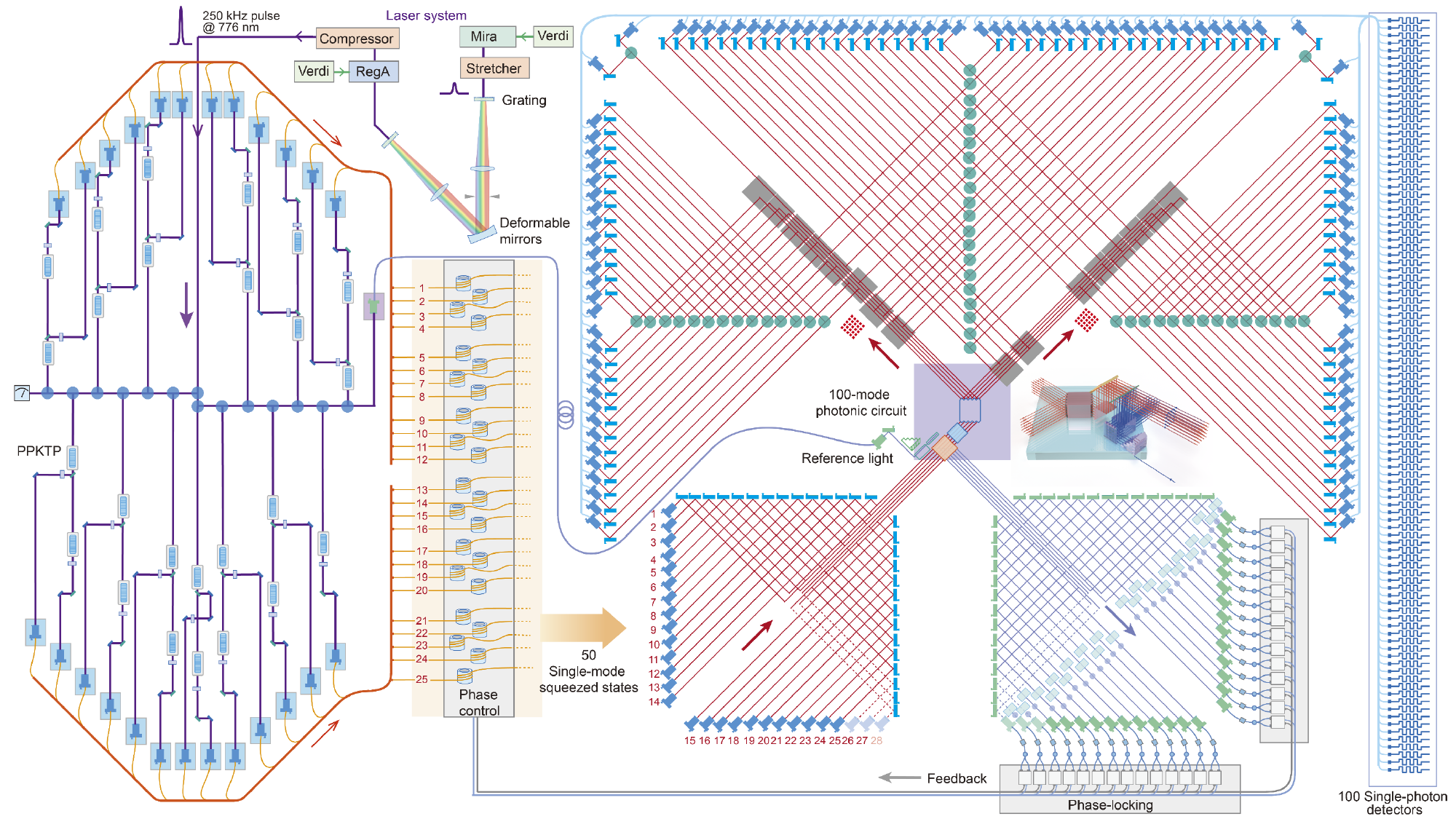} 
    \caption{\textbf{The \textit{Jiuzhang} light-based quantum computing device's configuration, engineered by USTC.} The machine operates by intricately manipulating light through an array of optical components. This visual representation of the \textit{Jiuzhang} photonic network provides insight into its experimental configuration, which occupies an optical table spanning an area of approximately three square meters. Within this setup, 25 Two-Mode Squeezed States (TMSSs) are introduced into the photonic network, resulting in the acquisition of 25 phase-locked light signals. To provide further elucidation, the output modes of the \textit{Jiuzhang} photonic network are systematically segregated into 100 distinct spatial modes through the employment of miniature mirrors and Polarizing Beam Splitters (PBSs). This accomplishment signifies the emergence of the second quantum computing system to assert the achievement of quantum computational advantages, following in the footsteps of Google's Sycamore quantum processor. Reproduced under the provisions of the Creative Commons license (\url{https://creativecommons.org/licenses/by/4.0/} from~\citep{Jiuzhang}).}
    \label{fig:JiuzhangNew}
\end{figure*}

\begin{figure*}
    \centering
    \includegraphics[width=\textwidth]{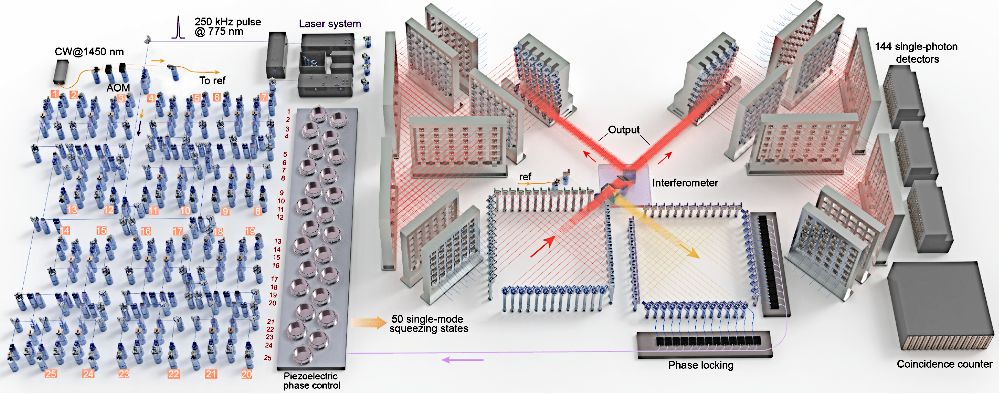}
    \caption{
\textbf{The \textit{Jiuzhang 2.0} experiment's configuration.} The 
experiment employs a configuration comprising five key elements. In the top-left quadrant, a high-intensity pulsed laser emitting light at a wavelength of 775 nm is utilized to excite 25 sources of TMSS (Two-mode Squeezed States), as indicated by the orange label within the left section. Simultaneously, a continuous-wave laser operating at 1450 nm is directed to co-propagate with the aforementioned 25 TMSS sources. The resulting 1550-nm two-mode squeezed light is conveyed into a single-mode fiber that demonstrates resistance to temperature variations. Notably, a 5-meter segment of this fiber is wound around a piezo-electric cylinder to enable control over the source phase, located in the central region. Transitioning to the central-right section, an optical arrangement comprising collimators and mirrors facilitates the injection of the 25 TMSSs into a photonic network. Here, 25 light beams corresponding to these TMSSs (illustrated in yellow) with a wavelength of 1450 nm and an intensity power of approximately 0.5 $\mu$W are harnessed for the purpose of achieving phase coherence. The resulting 144 output modes from this arrangement are divided into four segments using arrays of adjustable periscopes and mirrors. Ultimately, these output modes undergo detection using 144 superconducting nanowire single-photon detectors and are subsequently processed through a 144-channel high-speed electronic coincidence unit. 
Reproduced under Creative Commons license (\url{https://creativecommons.org/licenses/by/4.0/}) from~\citep{Jiuzhang2.0}.}
    \label{fig:Jiuzhang}
\end{figure*}

\subsection{Overview}

The USTC (University of Science and Technology of China) has made remarkable strides in photonic quantum computing technology with its series of \textit{Jiuzhang} quantum computers: \textit{Jiuzhang}~\citep{Jiuzhang}, \textit{Jiuzhang} 2.0~\citep{Jiuzhang2.0} and \textit{Jiuzhang} 3.0~\citep{Jiuzhang3.0}. Each iteration represents a significant milestone in the field. 
\textit{Jiuzhang} was the first to achieve quantum computational advantage~\citep{Jiuzhang}, marking a pivotal moment in quantum computing history. Subsequent versions, \textit{Jiuzhang} 2.0 and \textit{Jiuzhang} 3.0, have further advanced the field by enhancing programmability and speed. Notably, they have demonstrated remarkable capabilities in executing large-scale Gaussian boson sampling (GBS) tasks~\citep{GBS0,GBS,XanSoftw22,GBS1}, and solving nonplanar graph problems by leveraging GBS~\citep{GraphGBS}. 
As research and development in quantum computing progress, further advancements and improvements in these systems are expected to continue shaping the landscape of photonic QIP, paving the way for transformative applications in various fields.

\subsection{The USTC \textit{Jiuzhang}}

On December 3, 2020, USTC reported a significant achievement in the field of quantum computing~\citep{Jiuzhang}. Their quantum computer, \textit{Jiuzhang}, successfully executed GBS~\citep{GBS0,GBS,XanSoftw22,GBS1} within a mere 200 seconds. In contrast, the USTC research group estimated that the \textit{Sunway TaihuLight} supercomputer~\citep{Sunway} would require a staggering 2.5 billion years to perform the same calculation. 
The photonic quantum computer, \textit{Jiuzhang}, produces as many as $76$ photon clicks in its outputs~\citep{Jiuzhang}, resulting in an expansive state-space dimension of $10^{30}$. Notably, the sampling rate achieved by \textit{Jiuzhang} claimed to surpasses the capabilities of state-of-the-art simulation strategies and supercomputers by an astonishing factor of approximately $10^{14}$.

The experimental configuration consists of a Verdi-pumped Mira 900 Ti:sapphire laser (also known as titanium-sapphire lasers, Ti:Al$_2$O$_3$ lasers) which is divided into 13 paths of equal intensity. The laser beams are directed onto 25 periodically poled potassium titanyl phosphate (PPKTP) crystals, resulting in the generation of 25 two-mode squeezed states. A hybrid encoding technique is employed, effectively representing 50 single-mode squeezed states. To enhance the purity of the states, a 12nm filtering process raises the purity level from 98\% to 99\%. The 50 single-mode squeezed states are then fed into a 100-mode interferometer and subsequently detected by 100 single-photon detectors, each operating at an efficiency of 81\%.

Furthermore,  in order to draw a comparison between the results achieved by the photonic quantum computing and classical computing systems, an analysis was carried out in~\citep{Jiuzhang} to determine the time required for two supercomputers to execute an equivalent GBS task. The \textit{Taihu-Light} supercomputer was estimated to require approximately $8 \times 10^{16}$ seconds (equivalent to 2.5 billion years), whereas \textit{Fugaku} was projected to necessitate $2 \times 10^{16}$ seconds (equivalent to 0.6 billion years~\citep{Jiuzhang}. A schematic diagram illustrates the configuration of the light-based quantum computer \textit{Jiuzhang} is presented in Figure~\ref{fig:JiuzhangNew}.

\subsection{The USTC \textit{Jiuzhang} 2.0}

Subsequently, in the year 2021, an enhanced photonic quantum computer named ``\textit{Jiuzhang} 2.0,"  purposefully engineered for large-scale GBS was introduced in~\citep{Jiuzhang2.0}. The experiments involved a 144-mode photonic circuit, which yielded up to 113 photon detection events. The team achieved this feat through the development of a novel and scalable quantum light source based on stimulated emission of squeezed photons, boasting both near-unity purity and efficiency.

Remarkably, \textit{Jiuzhang} 2.0's photonic quantum computing capabilities translate into a substantial Hilbert space dimension of approximately $10^{43}$, leading to a sampling rate that is approximately $10^{24}$ times faster than brute-force simulations conducted on supercomputers. These developments represent a significant advancement in photonic quantum computing, paving the way for groundbreaking applications and furthering our understanding of quantum computational power.

The concept involves the generation of spontaneously produced photon pairs, resonating with the pump laser, to stimulate the emission of a second photon pair in a gain medium~\citep{jiuz25}. The experiment utilizes transform-limited laser pulses at a wavelength of 775 nm focused on PPKTP crystals to create two-mode squeezed states (TMSS). A concave mirror reflects the pump laser and the collinear TMSS photons back, serving as seeds for the second parametric process. Birefringence walk-off between horizontally and vertically polarized photons of the TMSS is compensated using a quarter-wave plate. Additionally, the dispersion between the pump laser and the TMSS is addressed by designing the PPKTP crystals to eliminate frequency correlation. Visual depictions of the experimental configurations are presented in Figure~\ref{fig:Jiuzhang}.

In their research, they achieved high collection efficiency (0.918 at a waist of 125 $\mu$m and 0.864 at 65 $\mu$m) and simultaneous high photon purity (0.961 and 0.946) without any narrowband filtering. Their double-pass approach can be readily expanded to higher orders, enabling the generation of higher brightness quantum light sources, scalable and nearly optimal for various applications.

\subsection{The USTC \textit{Jiuzhang} 3.0}

\begin{figure*}
    \centering
    \includegraphics[width=0.95\textwidth]{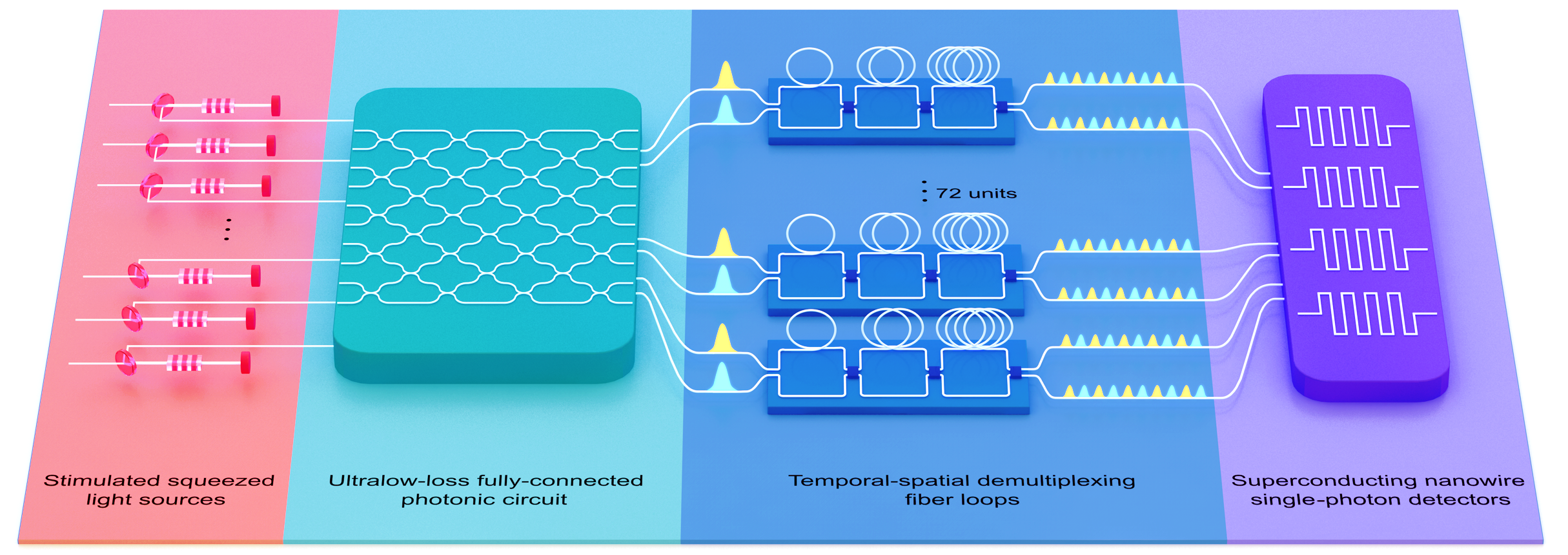}
    \caption{\textbf{\textit{Jiuzhang 3.0}: }
The experimental arrangement comprises 25 stimulated two-mode squeezed state photon sources, synchronized in phase, and directed into a 144-mode ultralow-loss fully-connected optical interferometer. These photons traverse 72 units of fiber loop setups for temporal-spatial demultiplexing before reaching 144 superconducting nanowire single-photon detectors, constituting a pseudo-photon-number resolving detection scheme. Each fiber loop setup, distinguished by distinct colors, accommodates two input modes. Within these setups, photons undergo temporal demultiplexing via fiber beam splitters and delay lines, resulting in four time bins. Furthermore, each time bin is partitioned into two path bins at the terminal fiber beam splitter. Discrimination between photons from the two input modes within the same fiber loop setup is facilitated by assessing their temporal parity through a coincidence event analyzer (not depicted).
Reproduced from ~\citep{Jiuzhang3.0} under a Creative Commons licenses (\url{https://creativecommons.org/licenses/by/4.0}).
}
    \label{fig:Jiuzhang 3.0}
\end{figure*}

In December 2023, a recent experiment in GBS utilizing pseudo-photon-number-resolving detection was detailed in~\citep{Jiuzhang3.0}. The experiment documenting photon-click events reaching a maximum of 255, see Figure~\ref{fig:Jiuzhang 3.0}. This investigation integrates considerations for partial photon distinguishability and introduces an extensive model for characterizing noisy GBS. Bayesian tests and correlation function analysis are applied within the context of quantum computational advantage to validate the samples against established classical spoofing simulations. Comparative assessments with state-of-the-art classical algorithms reveal that generating a single ideal sample from the identical distribution on the supercomputer \textit{Frontier} would demand approximately 600 years using exact methods, whereas the quantum computer, \textit{Jiuzhang} 3.0, achieves this feat in merely 1.27{$\mu$s}. Computing the most challenging sample from the experiment utilizing an exact algorithm would necessitate \textit{Frontier} approximately $3.1 \times 10^{10}$ years.

\section{ORCA Computing}

\subsection{Overview}

ORCA Computing~\citep{ORCA} is a leading company focused on the development of comprehensive photonic quantum computers. 
At its core, ORCA's vision encompasses the development of both near-term, value-generating quantum accelerators and long-term, error-corrected quantum computing systems, all made possible through its innovative modular, fiber-interconnected architecture. 
Since its founding in 2019, the company has successfully deployed its PT-1 quantum photonic system to clients worldwide, earning the trust of notable customers such as the UK Ministry of Defence (MoD) as they embark on their quantum computing journeys. The company's innovative approach harmoniously integrates state-of-the-art technology with readily available telecom and optical fiber components, promising to deliver unparalleled performance, scalability, and user-friendliness across a wide range of applications, including generative machine learning and optimization.

\subsection{ORCA's \textit{PT series} photonic quantum systems}

ORCA's quantum computing vision is driven by their innovative modular, fiber-interconnected architectures. 
They have successfully minimized component complexity by leveraging multiplexing and quantum memory technologies, thereby optimizing the quantum computing process. 
Importantly, ORCA leverages standard telecommunications technologies, enhancing robustness and cost-effectiveness. 
The modularity of ORCA's design not only supports scalability but also future upgradability, minimizing obsolescence and promoting ongoing research and development. 
Additionally, the \textit{PT series} by ORCA, designed for integration into classical computing workflows, is facilitated by ORCA's SDK, which is programmed in \textit{Python} and integrated with \textit{PyTorch}, simplifying access for machine learning users and fostering the development of quantum applications for practical use. This holistic approach caters to the hybrid quantum-classical environment and positions ORCA as a significant contributor to the advancement of quantum computing technologies in both the short and long term~\citep{ORCAfltlrnt,orcaAd2,orcaAd1}.

The PT Series by ORCA is specifically designed to integrate seamlessly with classical computing infrastructure and workflows. Furthermore, ORCA's Software Development Kit (SDK) is programmed in \textit{Python} and integrated with \textit{PyTorch}. This integration enables easy accessibility for machine learning users, enhancing productivity and reducing the complexity of developing AI applications for quantum computing.

\subsection{Fault tolerance in photonic MBQC architectures}

Fault-tolerant quantum computation (FTQC) hinges on the effective correction of hardware errors during program execution. The practical realization of fault tolerance is contingent upon the specific characteristics of the underlying hardware. For example, circuit-based error correction serves as a framework suitable for hardware featuring deterministic gates, allowing the detection of errors through non-destructive ancilla-assisted measurements~\citep{ORCAfltlrnt1,ORCAfltlrnt2}. Conversely, in the context of measurement-based quantum computing (MBQC), error syndromes are constructed through destructive measurements on previously generated entangled states. MBQC is notably well-suited for hardware involving probabilistic entangling operations and destructive measurements~\citep{ORCAfltlrnt3,ORCAfltlrnt4,ORCAfltlrnt5,PsiFBQC}, including discrete variable photonic qubits~\citep{PsiFBQC,ORCAfltlrnt7,ORCAfltlrnt8,ORCAfltlrnt9} and continuous variable qubits~\citep{ORCAfltlrnt10,XanArch6,ORCAfltlrnt12}.

Numerous photonic MBQC architectures~\citep{ORCAfltlrnt4,ORCAfltlrnt5,ORCAfltlrnt7,ORCAfltlrnt8,ORCAfltlrnt9,ORCAfltlrnt13,ORCAfltlrnt14,ORCAfltlrnt15,ORCAfltlrnt16} follow a two-stage approach to achieve fault tolerance. \textit{The first stage} involves creating a sizable entangled resource state whose dimensions correspond to the complexity of the quantum computer program. Subsequently, in \textit{the second stage}, destructive single-qubit measurements are applied to the prepared state to execute the program. A more streamlined approach in MBQC, referred to as fusion-based quantum computation (FBQC), was introduced in~\citep{PsiFBQC}. In FBQC, destructive two-qubit projective measurements in the Bell-state basis, known as Bell-state measurements~\citep{PhoInformProc224} (BSMs) or fusions~\citep{ORCAfltlrnt18}, are performed on constant-sized resource states. The study in~\citep{PsiFBQC} also presents FBQC architectures implementing the surface code, demonstrating high thresholds for both photon loss and fusion failures. Similar high thresholds have been achieved for other topological error correction codes using FBQC~\citep{ORCAfltlrnt19,ORCAfltlrnt20}. The ability to generate entangled photonic resource states~\citep{ORCAfltlrnt21,ORCAfltlrnt22,ORCAfltlrnt23,ORCAfltlrnt24,Ascella}, along with the high thresholds, suggests the suitability of photonic platforms for implementing FBQC architectures. However, the resource state size and photon loss thresholds remain challenging for current hardware.

In~\citep{ORCAfltlrnt}, ORCA Computing has explored alternative photonic architectures aimed at mitigating the hardware demands. Specifically, they have devised MBQC architectures that attain fault tolerance by fusing n resource states through measurements in the n-qubit Greenberger–Horne–Zeilinger (GHZ) state basis, known as GHZ-state measurements (GSMs).

\subsection{ORCA's path to fault tolerance}

Recently~\citep{ORCAfltlrnt}, ORCA Computing introduced innovative fault-tolerant architectures.
Their focus lies on utilizing projective measurements in the GHZ basis in conjunction with constant-sized entangled resource states. They have meticulously developed linear-optical constructions of these architectures, wherein the GHZ-state measurements are subjected to encoding techniques aimed at mitigating errors stemming from photon loss and the inherent probabilistic characteristics of linear optics. Extensive simulations within their study indicate remarkable enhancements in single-photon loss thresholds when compared to the current state-of-the-art linear-optical architectures, specifically those employing encoded two-qubit fusion measurements with constant-sized resource states. ORCA Computing's findings in this endeavor underscore the potential of a resource-efficient route towards achieving fault-tolerant quantum computing based on photonic technologies.

An extensive array of research directions, extending beyond the scope of ORCA Computing's work presented in~\citep{ORCAfltlrnt}, encompasses various avenues. These include the investigation of alternative coding schemes for GHZ-state measurements as delineated in~\citep{ORCAfltlrnt45,ORCAfltlrnt46}, the exploration of leveraging biased error correction techniques as discussed in~\citep{ORCAfltlrnt19}, the utilization of diverse resource states as highlighted in~\citep{ORCAfltlrnt20}, and the examination of fusion network designs not rooted in conventional foliation principles as discussed in~\citep{ORCAfltlrnt47,ORCAfltlrnt48}.

Beyond the realm of quantum computation, the encoded GHZ-state measurements developed by ORCA may also find pertinent applications in the domain of quantum communication. For instance, ORCA Computing exhibit potential utility in facilitating multipath routing strategies, which, in turn, hold the promise of enhancing entanglement rates within quantum networks, as elucidated in~\citep{ORCAfltlrnt49,ORCAfltlrnt50}. 
Photonic resource state generation methods can be found in Refs.~\citep{ORCAfltlrnt9,ORCAfltlrnt18,ORCAfltlrnt23,ORCAfltlrnt26,ORCAfltlrnt27,ORCAfltlrnt28}.

\section{Photonic}

\subsection{Overview}

Photonic~\citep{Photonic} is in the process of constructing an innovative quantum computing and networking platform, distinguished by its scalability, fault-tolerance, and integration~\citep{PhotonicJourney}. Founded in 2016, the company harnesses a proprietary technological framework encompassing photon spin interfaces within silicon, integrated silicon photonics, and quantum optical components~\citep{PhotonicTcentres,PhotonicSilicon,PhtncFault25,PhtncFault26}. Photonic's technological framework includes a native telecommunications networking interface and leverages the manufacturing capabilities inherent in silicon~\citep{PhtncFault78,PhtncFault79,PhotonicNetwSupercom}. Photonic collaborating with Microsoft to power global quantum ecosystem~\citep{PhotonicMicrosoft}.

\subsection{Silicon spin-photon interface}

Photonic's quantum computing approach is distinctively characterized by the integration of silicon spin qubits with photons, in order to surmounting the obstacles that hinder the progress of other quantum computing frameworks~\citep{PhotonicTcentres}. This integrated solution leverages photonically-linked silicon spin qubits, facilitating the realization of a resource-efficient, error-corrected, and genuinely scalable quantum computing platform~\citep{PhotonicFault,PhtncFault145}. The architecture, known as the silicon spin-photon platform, paves the way for the achievement of scalable fault-tolerance~\citep{PhotonicFault}. By harnessing the computational potential of spin qubits with photon interfaces, this architectural design enables seamless communication through ultralow-loss telecommunication band fibers~\citep{PhtncFault49,PhtncFault124,PhtncFault133}, thereby enabling fault tolerance at a large scale~\citep{PhotonicNetwSupercom}. Furthermore, this architecture amalgamates the advanced manufacturing capabilities of silicon with distributed computing, ultimately yielding a quantum computing system with boundless scalability potential~\citep{PhotonicFault}.

\subsection{Photonic's path towards fault-tolerant quantum technologies}

\begin{figure*}
    \centering
    \includegraphics[width=\textwidth]{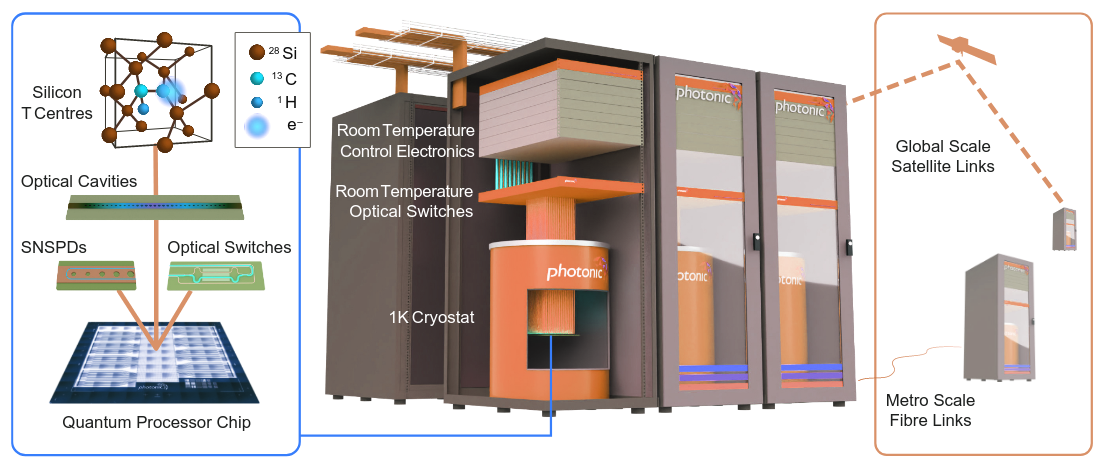}
    \caption{
    \textbf{Photonic's scalable quantum technology architecture. } Involves a quantum chip that undergoes cooling within a 1 K cryostat. This chip incorporates integrated silicon T centers situated within optical cavities, along with photonic switches and single-photon detectors. The architecture includes optical input-output (IO) ports connected to a room temperature photonic switch network and control electronics through telecom fiber. This design naturally facilitates a highly-connected architecture characterized by non-local connectivity, even as the system scales in size. The utilization of telecom fiber enables horizontal scaling of the system by interconnecting multiple cryostats through their optical IO. This, in turn, enables the expansion of computing power and the establishment of long-distance quantum networks. 
    Reproduced with alternations from (\url{www.photonic.com}).}
    \label{fig:photonic_arch}
\end{figure*}

The existing challenges in scaling quantum networking and quantum computing technologies share a common obstacle of achieving widespread distribution of high-quality entanglement. In this regard,  Photonic proposed a novel QIP architecture, centered on optically active spins in silicon~\citep{PhotonicFault,PhtncFault25,PhtncFault26,PhtncFault27,PhtncFault28}. This architecture (see Figure~\ref{fig:photonic_arch}) presents an integrated technological platform that addresses the requirements for scalable, fault-tolerant quantum computing and networking~\citep{PhotonicFault}. Emphasizing optimization for overall entanglement distribution, the design utilizes color center spins in silicon (T centres)~\citep{PhtncFault25,PhtncFault98,PhtncFault99} for their manufacturability, photonic interface, and high-fidelity information processing attributes~\citep{PhotonicFault}.

\begin{figure*}
    \centering
    \includegraphics[width=\textwidth]{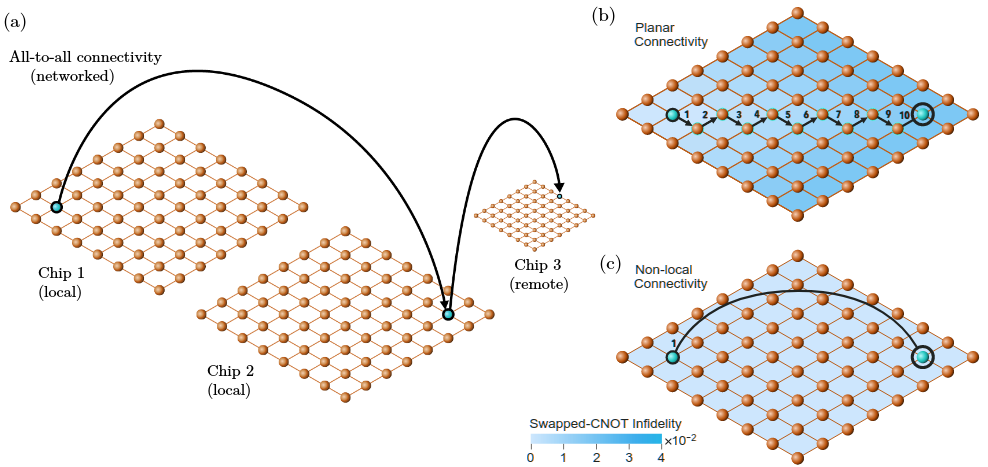}
    \caption{ T center network. 
    (a) The non-local connectivity of the system operates consistently, whether entangling qubits within the same chip or extending across multiple chips in the same cryostat, spanning distances of 10 meters or even 100 kilometers. This underscores the robustness and scalability of the system, showcasing its ability to maintain performance integrity irrespective of spatial separation. 
    The comparison between two-qubit operations on devices featuring: 
    (b) planar, nearest-neighbor connectivity graphs and (c) non-local connectivity reveals distinct efficiency disparities. In the case of a planar graph, executing a CNOT gate between two distant qubits necessitates sequential swap operations. Each successive operation introduces errors, resulting in a significant decline in overall fidelity, even over relatively short distances depicted in the illustration. Conversely, non-local connectivity ensures that two-qubit operations remain uniform across the entire graph, allowing the device to scale with enhanced fidelity.
    Adapted from~\citep{PhotonicNetwSupercom}.}
    \label{fig:photonic_comp}
\end{figure*}

This architectural framework is broadly applicable across various qubit systems~\citep{PhotonicFault}. The spin-photon interface's elevated connectivity allows for the utilization of fixed- and low-overhead quantum low-density parity-check (QLDPC) codes~\citep{PhtncFault48,PhtncFault49,PhtncFault50,PhtncFault51,PhtncFault52,PhtncFault53,PhtncFault72,PhtncFault55,PhtncFault56,PhtncFault57,PhtncFault58,PhtncFault59,PhtncFault60,PhtncFault61,PhtncFault62,PhtncFault63,PhtncFault64,PhtncFault65,PhtncFault66,PhtncFault67,PhtncFault68,PhtncFault69,PhtncFault70,PhtncFault71,PhtncFault72}, ensuring fault tolerance. Leveraging an integrated silicon photonics platform, thousands of qubits can be manufactured and addressed on a single chip, incorporating both optical and electronic control and routing~\citep{PhotonicTcentres,PhotonicSilicon,PhtncFault126,PhtncFault127,PhtncFault128,PhtncFault129,PhtncFault130}. Furthermore, modules can be seamlessly interconnected through existing global telecommunications infrastructure without incurring transduction losses. Employing a T center  network~\citep{PhotonicTcentres,PhtncFault25,PhtncFault98,PhtncFault99} (as shwon in Figure~\ref{fig:photonic_comp}) for distributing verified quantum entanglement enables device-independent networking protocols, offering robust protection against eavesdropper attacks. This approach also has significant implications for applications dependent on entanglement distribution, such as blind computing~\citep{PhtncFault15,PhtncFault97}.

\section{PsiQuantum}

\subsection{Overview}

PsiQuantum~\citep{PsiQuantum} employs a photon-based approach to qubit production, wherein photons traverse silicon chip-embedded conduits. By utilizing mirrors, these photons are maneuvered into an entangled state, and their fusion measurements serve as gates. 
The company, founded in 2015, asserts that their approach represents the sole practical method for constructing a commercially viable quantum computer with a requisite number of qubits. Their goal is to accelerate the development of dependable systems 
leveraging existing manufacturing processes and infrastructure.

PsiQuantum adopts a superconducting single-photon detector, renowned for its capacity to achieve requisite efficiencies with minimal developmental  efforts. These detectors operate at approximately 4 Kelvin, a temperature notably higher than the milli-Kelvin temperatures required for superconducting qubits~\citep{SC-qubits} and certain competing technologies~\citep{DiVincenzo}.

\subsection{PsiQuantum’s approach to fault-tolerance}

\textit{Fusion-based quantum computation }
The concept of FBQC has been introduced as a novel approach to quantum computing architecture~\citep{PsiFBQC}. Traditional quantum computing relies on deterministic unitary entangling gates~\citep{Entangling,SQSCZ1,SQSCZ2}, which may not align naturally with certain physical systems, notably those in the realm of photonics~\citep{PsiFBQC1,PsiFBQC2,ORCAfltlrnt18,PsiFBQC4,PsiFBQC5,PsiFBQC6,ORCAfltlrnt7,ORCAfltlrnt8,ORCAfltlrnt16,Deterentang,Deterentang1,Deterentang2,Deterentang3,Deterentang4,PsiFBQC2,Deterentang6}. In contrast, FBQC leverages physical primitives inherently accessible in photonic setups, primarily entangling measurements termed ``fusions," performed on qubits originating from small, constant-sized entangled resource states. This approach not only addresses the probabilistic nature of photonic gates and inherent errors but also integrates quantum error correction techniques effectively~\citep{PsiFBQC}.

PsiQuantum's claims that this computational model can attain significantly higher fault tolerance thresholds than those documented in existing literature~\citep{PsiFBQC}. Specifically, the company proposes a robust design capable of withstanding a substantial $10.4\%$ probability of photon loss during each fusion, equating to a mere $2.7\%$ chance of losing each individual photon~\citep{PsiFBQC}.  PsiQuantum also demonstrates a $43.2\%$ barrier against fusion failure using a ballistic technique~\citep{PsiFBQC10}, compared to $14.9\%$ previously reported in~\citep{ORCAfltlrnt5}. Notably, this architecture offers modularity and substantially reduces classical processing demands compared to earlier photonic quantum computing designs~\citep{ORCAfltlrnt8,ORCAfltlrnt16,PsiFBQC10,ORCAfltlrnt5}.

\begin{figure*}
    \centering
    \includegraphics[width=0.97\textwidth]{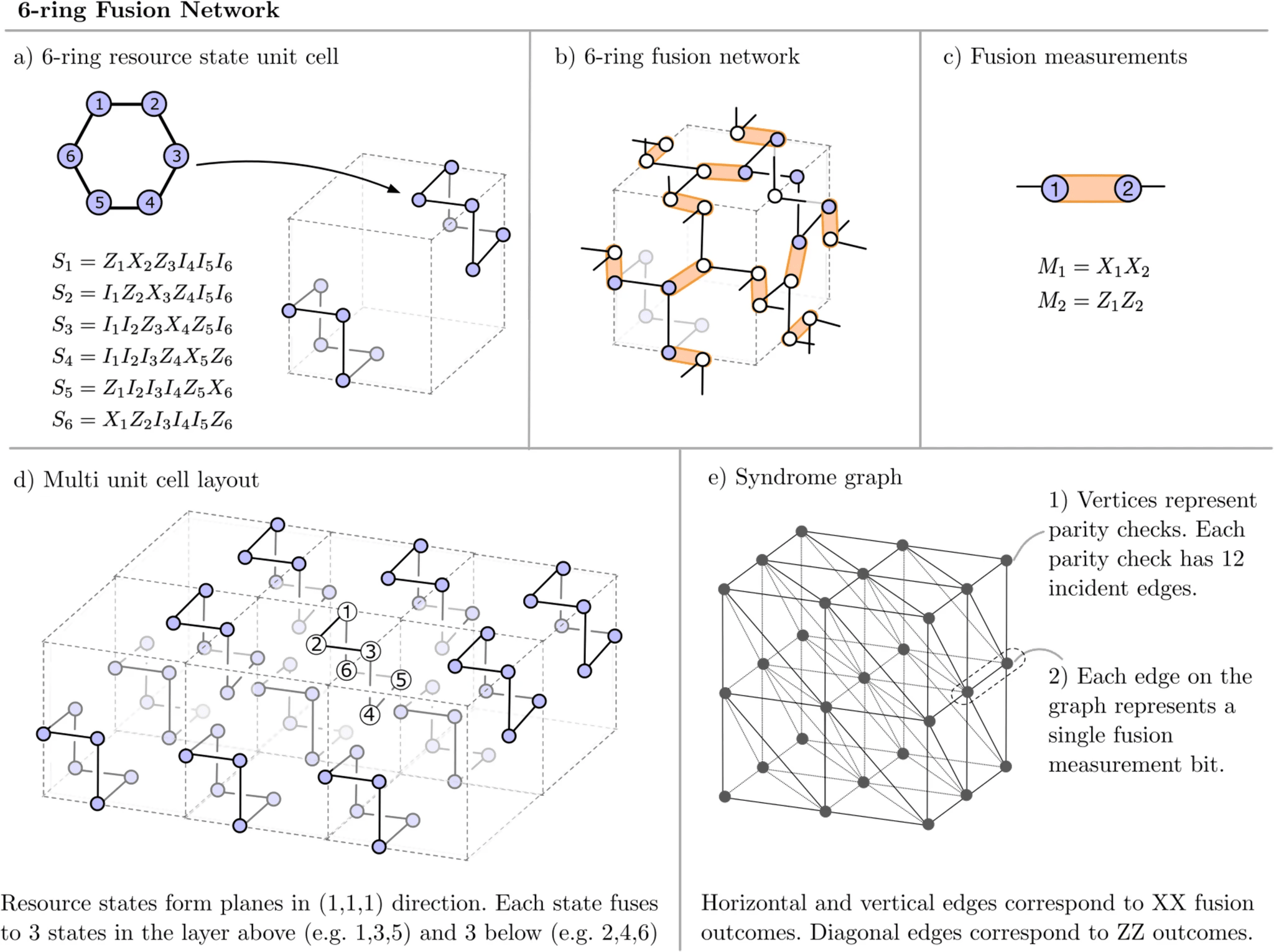}
    \caption{\textbf{The ``Six-ring" Fusion Network.}
(a) Each resource state is represented as a graph state configured in the shape of a ring containing six qubits. Two resource states are positioned at opposite corners of each unit cell.
(b) Two-qubit fusions establish connections between every pair of qubits that share a common face or edge. Resource states belonging to the unit cell are denoted as purple circles, while qubits from resource states in neighboring cells are depicted as white circles. 
(c) All fusion measurements within the fusion network are two-qubit projective measurements conducted in the bases $M_1=X_1 X_2$ and $M_2=Z_1Z_2$.
(d) Depicts the arrangement of resource states spanning multiple unit cells. When these unit cells are arranged in a tessellated manner, the resource states can be organized into layers along two-dimensional planes perpendicular to the (1,1,1) direction. Each state involves the fusion of three qubits with the layer above and three with the layer below.
(e) The syndrome graph arising from the fusion layout exhibits a cubic graph structure with diagonal edges, as shown. Primal and dual syndrome graphs share an identical configuration. In both, the vertical edges correspond to XX-type fusion outcomes, while diagonal edges correspond to ZZ outcomes. The unit cells of the primal and dual syndrome graphs can be interpreted as shifted by (1/2,1/2,1/2) so that each fusion event corresponds to both a primal and dual edge, intersecting perpendicularly at the fusion location itself. 
Reproduce from~\citep{PsiFBQC} under a Creative Commons Attribution 4.0 International License 
(\url{http://creativecommons.org/licenses/by/4.0/}).}
    \label{fig:6-ring}
\end{figure*}

\textit{Flexible architectural design in fusion networks } 
The physical architecture of FBQC, termed fusion networks, features a remarkable degree of flexibility, devoid of inherent temporal or spatial constraints. This flexibility allows for variations in the physical implementation of fusion networks. 
At the core of this architecture lies the concept of Resource State Generators (RSGs), physical devices responsible for generating resource states at specific spatial locations and time intervals. This concept is especially pertinent to photonic architectures~\citep{Photonics1}. The lifespan of a qubit is a critical factor to consider; it commences with its creation within an RSG, then proceeds through a fusion network router, which directs the qubits to their respective fusion locations~\citep{PsiFBQC}. Ultimately, each qubit undergoes a destructive measurement within a fusion. This relatively brief qubit lifespan proves advantageous for FBQC, especially within photonic architectures where optical loss remains a dominant source of physical errors~\citep{PsiFBQC31,PsiFBQC32,PsiFBQC33}. Figure~\ref{fig:6-ring} illustrates a 6-ring example fusion network. 
In~\citep{PsiFBQC} PsiQuantum introduces FBQC as a universal quantum computation model that capitalizes on the generation of small, constant-sized entangled resource states and projective entangling measurements. 
The research explores how this model can be harnessed to achieve topological fault-tolerant quantum computation, particularly in the context of photonic architectures~\citep{PsiFBQC23,PsiFBQC24,PsiFBQC25,PsiFBQC26,PsiFBQC27,PsiFBQC28,PsiFBQC29,PsiQ1}.

\textit{Towards large-scale, fault-tolerant quantum computation with FBQC } 
FBQC is formulated as a versatile framework for quantum computation, aligning with the inherent characteristics of various physical systems, including those in the domain of photonics. In~\citep{PsiFBQC} the authors anticipate that this framework, which tightly integrates physical errors with their impact on quantum error correction, holds the potential to enhance the performance of systems fundamentally based on resource state generation and projective measurements, as exemplified in linear optical quantum computing (LOQC)\citep{ORCAfltlrnt8,PsiFBQC10,ORCAfltlrnt5,PsiFBQC68,PsiFBQC69,PsiFBQC70,ORCAfltlrnt15,PsiFBQC2,PsiFBQC4}. PsiQuantum's research culminates in the demonstration of a remarkable doubling of fault tolerance thresholds compared to previous schemes. As quantum technology advances closer to realizing these systems, the theoretical framework introduced by PsiQuantum assumes significance as a vital tool for engineering both hardware and architectural designs, facilitating the realization of large-scale, fault-tolerant quantum computation. 
Further elaboration on this physical architecture is provided in~\citep{PsiFBQC}.

\subsection{PsiQuantum’s active volume architecture}

PsiQuantum's research in~\citep{PsiQ3}, introduces a decoding framework to support FTQC. 
In their pursuit, they introduced the Active Volume Architecture (AVA)~\citep{PsiQ5}, a technique expected to amplify quantum algorithm efficiency by approximately 50 times, specifically targeting error-corrected quantum systems. This method optimizes hardware utilization by facilitating long-range connections within the quantum computer, particularly advantageous in photonic quantum computing. 
Their advancements in FTQC extend to impactful domains like cryptography~\citep{PsiQ4}, estimating the computational scale required to breach widely used cryptographic systems using a novel architecture, leveraging non-local connections from photonic components. Collaborating with Boehringer Ingelheim~\citep{Ingelheim}, PsiQuantum explores molecular observable computations, critical for industrial applications with quantum computers.

In partnership with Mercedes-Benz~\citep{PsiMercedes}, PsiQuantum investigates the quantum computational thresholds necessary to innovate Li-ion battery design, leveraging their expertise in Fault Tolerance. Additionally, extensive research on photonic quantum computing technologies by PsiQuantum is detailed in~\citep{PsiMercedes,PsiQ1,PsiQ2,PsiQ3,PsiQ4,PsiQ5,PsiQ1,PsiQ7,PsiQ8,PsiQ9,PsiQ10,PsiQ11,Univ12-mod12,PsiQ13,Univ12-mod1,PsiQ15,Univ12-mod2,PsiQ17,PsiQ18,PsiQ19,PsiQ20,PsiQ21,PsiMercedes,ORCAfltlrnt7,PsiQ24,ORCAfltlrnt28,ORCAfltlrnt7,PsiQ27,PsiQ28}.

\section{Quandela Photonic Quantum Computers}

\subsection{Overview}

Quandela is a pioneering quantum technology company that provides a comprehensive suite of solutions, encompassing quantum software, middleware, and hardware~\citep{QuanDevic1,QuanDevic2,QuanDevic3,QuanDevic4,QuanDevic5,QuanDevic6}, all firmly grounded in photonic technologies. 
Quandela, was founded in 2017. Quandela has achieved global recognition for its pioneering work in the development of solid-state quantum light sources. Notably, Quandela's \textit{Prometheus}, is touted as the world's first photonic qubit generator, serving diverse applications such as quantum cryptography, quantum computation, and quantum sensors~\citep{Prometheus}.

Quandela is dedicated to advancing quantum technology, exemplified by its introduction of a two-qubit quantum processor designed to enhance cybersecurity. Additionally, the company has introduced \textit{Perceval}~\citep{Perceval}, a photonic quantum computing software platform, as part of its commitment to providing a comprehensive suite of quantum solutions~\citep{Perceval,QuanAppl1,QuanAppl2,QuanAppl3,QuanAppl4,QuanAppl5}. These strategic innovations underline Quandela's 
role in exploring the full potential of solid-state quantum light emitters in the realm of large-scale quantum computing~\citep{QuanQCom1,OneNine,QuanQCom3,Ascella,QuanQCom5,randomness,QuanQCom7,QuanQCom8,QuanQCom9,QuanQCom10,QuanQCom11,QuanQCom12,QuanQCom13,QuanQCom15}.

\subsection{\textit{MosaiQ}}

Quandela's introducing photonic quantum computing platform, known as \textit{MosaiQ}, represents a cutting-edge solution for the precise manipulation of exceptionally pure and on-demand photons~\citep{MosaiQ}. This platform serves as a catalyst for the exploration, refinement, and deployment of an innovative generation of photonic quantum computing algorithms and protocols~\citep{QuanAppl1,QuanAppl2,QuanAppl3,QuanAppl4,QuanAppl5}. Distinguishing itself through a modular and expandable architecture, \textit{MosaiQ} seamlessly integrates Quandela's eDelight sources, active demultiplexing through DMX modules, and an integrated circuit-based photonic processor, which is not only fully reconfigurable but also operates efficiently at room temperature~\citep{quandela13}. Moreover, it incorporates top-tier nanowire detectors and rapid electronic time-tagging modules. With a range of \textit{MosaiQ} Quantum Processing Units (QPUs) at one's disposal, spanning from 2 to 12 photonic qubits~\citep{12mod,QuanCat}.

\textit{A Compact, Swift, and Efficient Quantum Device }
The QDMX-6~\citep{QuanCat}, a remarkable advancement in quantum technology, is characterized by its compact, rapid, and resource-efficient design. It serves as the inaugural active temporal-to-spatial demultiplexer tailored specifically for quantum applications, seamlessly merging optical and electronic components within a condensed module. Table~\ref{QDMX6} summarize the technical specifications and features of the QDMX-6. This innovative device is now readily available to accommodate the routing of up to six photons, meticulously crafted to interface with eDelight sources through integrated circuits.

\begin{table*}
\caption{\label{QDMX6}Technical specifications and features of the Quandela's QDMX-6 quantum device~\citep{QuanCat}.}
\centering
\begin{tabular}{p{6cm}|p{9cm}}
 \hline \hline
\textbf{Specification} & \textbf{Details} \\
\hline
Technology & Proprietary design for the integration of active elements in combination with optics for high-speed routing and low transmission losses. \\
\hline
Electrical driving & Output voltage signal (DC) \\
\hline
Number of independent outputs (single mode fiber outputs)& Up to 6 in a single module, with a tunable sequence.\\
\hline
Optical frequency $\eta$ &(Total optical transmission/line) \\& $>$75\% \\
\hline
Speed, duty cycle, and metrics\footnotemark[1] & $T_{\text{switch}}$ $\sim$ 50 ns\\
& $T_{\text{plateau}}$ $>$ 50 ns (typically 50 ns - 150 ns)  \\
&(tunable depending on the application)\\
\hline
Optical and mechanical stability & Limited signal fluctuation over days \\
\hline
Physical dimensions & Height: 150 mm\\
                    & Width: 490 mm\\
                    & Length: 470 mm\\
                    & Weight: $\sim$17 kg (optics); $\sim$4 kg (electronics)\footnotemark[2] \\
                    \hline
Electrical connections & 100V/120V/230V, 50Hz \\
\hline \hline
\end{tabular}
\footnotetext[1]{
From the provided metrics, it is possible to calculate the N-photon coincidence rate at the output of the QDMX-N device. To do this,  the``filling factor" have to be determined first, which serves as a reliable approximation of the effective events. The filling factor (${\cal FF}$) can be calculated as 
${\cal FF} = T_{\mathrm{plateau}}/(T_{\mathrm{switch}} + T_{\mathrm{plateau}})$. 
With the filling factor in hand, the final rate of $C_N$ coincidences at the output can be calculated. The formula for $C_N$ is given as: 
$C_N = {\cal R}  \times  {\cal FF} \times  \frac{(\eta \times {\cal B})^N}{N}$.   
Here, ${\cal R}$ represents the clock rate of the driving excitation laser, and ${\cal B}$ identifies the efficiency of the eDelight source. This equation allows to estimate the rate of N-photon coincidences based on the system's parameters and characteristics.}
\footnotetext[2]{Fiber delays not considered.}
\end{table*}

\subsection{\textit{Prometheus}}

\textit{Prometheus} is distinguished as the autonomous quantum light source, representing a significant advancement in the domain of quantum technologies~\citep{Prometheus}. It consolidates a cryostat, laser system, solid-state single-photon sources\citep{PsiFBQC1,ORCAfltlrnt8,siglPhoton14,ORCAfltlrnt5,PsiFBQC,quandela1,quandela2,quandela9}, and a qubit control unit into a single, compact system, positioning it as a pioneering development in the field~\citep{QuanCat}.

\textit{Prometheus}~\citep{Prometheus} benefits from Quandela's distinctive design and fiber-pigtailing technique, enabling the seamless integration of all essential opto-electronic, cryogenic, and solid-state quantum light sources into a 19-inch rack~\citep{Prometheus}. This feature enhances \textit{Prometheus}'s portability and adaptability, rendering it an invaluable tool for quantum experiments across different environments. The system produces high-quality photonic qubits at unmatched rates, simplifying access to the potential of single photons for both academic and industrial users, with straightforward operation.

\textit{Prometheus}~\citep{Prometheus} excels in terms of ease of use, performance, and reliability, essential attributes in the field of optical quantum technologies. It ensures a consistent stream of identical single photons generated in a stable and robust manner. The design of \textit{Prometheus} as a standalone single-photon source makes it an ideal solution for applications that necessitate a high rate of single and indistinguishable photons~\citep{quandela1,quandela2,quandela9}. The all-in-one device provides a steady photon stream with outstanding brightness~\citep{QuanDevic6}.

\begin{figure*}
    \centering
      \includegraphics[width=0.9\textwidth]{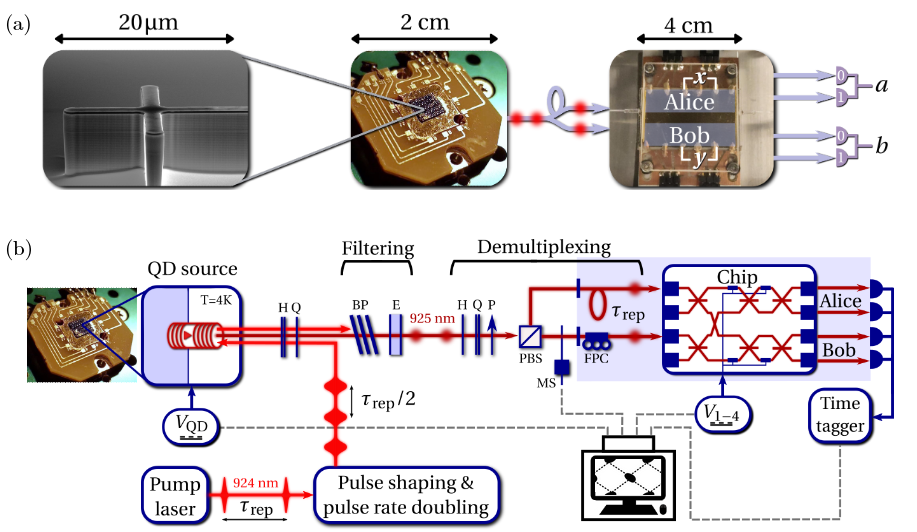}
    \caption{\textbf{Quandela’s two-qubit random number generator.}
    (a) A compact implementation of a certified quantum random number generator. The quantum dot device produces individual photons at a wavelength of 925 nm. These photons are synchronized using a fibered delay and sent into a glass chip, which generates a post-selected entangled Bell pair and allows for the implementation of measurement bases x and y for Alice and Bob. High-efficiency single photon detectors are employed to simultaneously detect photons by Alice and Bob, yielding a corresponding coincidence result $(a, b)$. 
(b) The diagram illustrates the experimental setup for the quantum random number generator. Photons at 925 nm are generated by a quantum dot (QD) photon emitter through a phonon-assisted excitation process. The setup includes optical components such as half and quarter-wave plates (H, Q), bandpass filters (BP), and an etalon (E). The polarizing beamsplitter (PBS) collects the outputs, and the entire setup is connected via fibers or waveguides within the blue-shaded area. A fibered delay ($\tau$ rep) ensures the synchronization of photon pairs sent into the chip. A motorized shutter (MS) enables the calibration of chip voltage. A fibered polarization controller (FPC) guarantees that both photons enter the photonic chip with matching polarization. Dashed gray lines indicate that certain components of the setup are automated to implement the randomness generation protocol, including adjustments to the voltage on the photon source to optimize brightness and periodic calibrations of thermo-optic phase-shifter voltages. The symbols $V_{1-4}$ control the phases on the chip, determining the measurement bases for Alice and Bob. A feedback loop represented by $V_{QD}$ ensures that the quantum dot emission remains bright, and the emitted photons remain indistinguishable. Reproduced under Creative Commons Attribution license (\url{https://creativecommons.org/licenses/by/4.0/}) from~\citep{randomness}.
    }
    \label{fig:numbgene}
\end{figure*}

Figure~\ref{fig:numbgene} illustrates the implementations of randomness generation protocols certified through nonlocality or contextuality~\citep{randomness}. The presented configuration marks the first on-chip certified randomness generation protocol, backed by a thorough security analysis that maintains its resilience against quantum side information. Its core objective is to successfully mitigate the locality loophole within a compact device. It's worth noting that additional protocols can be found in ~\citep{randomness3,randomness4,randomness5,randomness21,randomness31,randomness32,randomness52,randomness53,randomness54,randomness55}.

\begin{table*}
\caption{Specifications for Quandela's stand-alone quantum light source - \textit{Prometheus}. \textit{Prometheus} offers the capability to provide exceptional rates of high-quality photonic qubits, affording both academic and industrial sectors convenient access to the potential of single photons with remarkable ease~\citep{QuanCat}.}
\centering
\begin{tabular}{p{6cm}|p{8cm}}
 \hline \hline
\textbf{Specification} & \textbf{Details} \\
\hline
Technology & Proprietary design. \\
& Deterministic fabrication of the source devices and optical fiber pigtailing technique. \\
\hline
Emission Wavelength & 925 $\pm$ 5 nm \\
& 780 nm (available from 2022) \\
\hline 
Fibered Brightness\footnotemark[1] & $>$13\%  \\
(Polarized Single-Photon Emission Probability under Resonant Excitation) & (corresponding to a single-photon rate of 10.4 MHz for an 80 MHz laser repetition rate) \\
\hline 
Single Photon Purity ($g^{(2)}(0)$) & Typically 2-4\% \footnotemark[2] \\
\hline 
Photon Indistinguishability & $>$91\% \footnotemark[3]\\
& (for operation at 4K) \\
\hline
Single-Photon Bandwidth  & 1.2 $\pm$ 0.4 GHz  \\
(Emitter Lifetime)&$<$150 $\pm$ 50 picoseconds \\ 
&``Fourier-transform-limited" emission \\
\hline
Cool Down Time & 15 minutes (40K)  \\
& 4 hours ($<$4K) \\
\hline
Compressor & $<$4K (Air-cooled, Optional: Not racked water-cooled) \\
& 40K (Air-cooled)\\
\hline 
User Interface & Fully automated control of the different modules using the central computer\\
\hline 
Power Consumption and Electrical Connections & $<$3 kW,  220V AC for air-cooled\\
\hline
Physical Dimensions & Height: 180 cm (31RU available space) \\
                    & Width: 76 cm \\
                    & Depth: 108 cm \\
                    & Weight: 250 kg \\
\hline \hline
\end{tabular}
\footnotetext[1]{The enhancement of the single-photon emission rate can be achieved by adjusting the laser repetition rate, which can be accommodated upon request. Additionally, continuous wave excitation is feasible upon request as well.}
\footnotetext[2]{Quantification is conducted using the``second order correlation measurement $g^2$" method, employing a Hanbury Brown-Twiss interferometer. Where $g^{(2)}$ is the normalized
second-order correlation function.}
\footnotetext[3]{Measurement is performed utilizing 'Hong-Ou-Mandel' interference measurements.}
\end{table*}

\subsection{\textit{Ascella}}

Quandela introduces \textit{Ascella}~\citep{Ascella}, a pioneering, user-accessible, general-purpose quantum computing prototype leveraging single photons~\citep{PsiFBQC1,ORCAfltlrnt8,siglPhoton14,ORCAfltlrnt5,PsiFBQC}. \textit{Ascella} is a quantum computer that integrates a high-efficiency quantum-dot single-photon source~\citep{QuanDevic6,quandela1,quandela2,quandela9} to power a universal linear optical network on a reconfigurable chip~\citep{Ascella18}. The hardware imperfections are addressed through a transpilation process driven by machine learning~\citep{OneNine}.

The compensation for hardware errors involves the adaptation of input circuits to align with the specific quantum device topology. Additionally, \textit{Ascella} features a comprehensive software stack enabling remote control, facilitating diverse computational tasks via logic gates or direct photonic operations. Quandela's evaluation of one-, two-, and three-qubit gates for gate-based computation yielded exceptional fidelities of $99.6 \pm 0.1\%, 93.8 \pm 0.6\%$, and $86 \pm 1.2\%$ respectively~\citep{Ascella}.

\textit{Ascella} implement a variational quantum eigensolver (VQES)~\citep{PsiQ13,Ascella46,Ascella49}, demonstrating high-precision calculations of hydrogen molecule energy levels. Furthermore, it showcases a 3-photon-based quantum neural network~\citep{Ascella50,Ascella51,Ascella52,Ascella53,Ascella54} for photon-native computation and achieves a significant milestone in quantum computing with a first-ever demonstration of 6-photon Boson sampling on a universal reconfigurable integrated circuit~\citep{Ascella,QuanAppl2,Ascella61,Ascella62,Ascella63}.

\begin{figure*}
    \centering
   \includegraphics[width=0.9\textwidth]{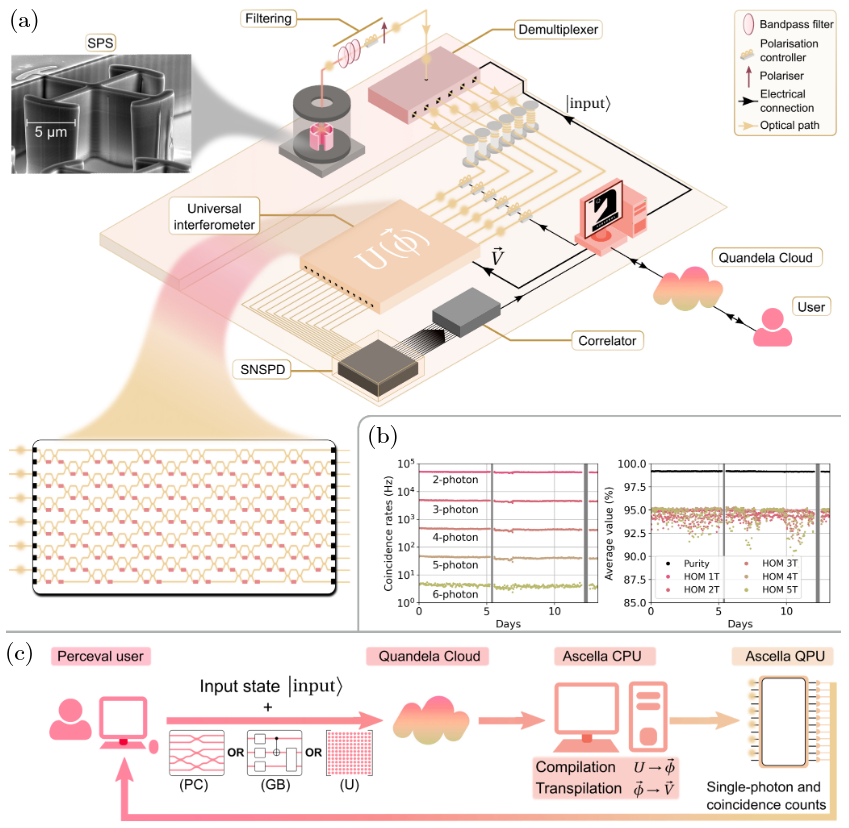}
    \caption{
    \textbf{Ascella's architecture, performance, and robustness.}
    (a) The comprehensive design of a 6-photon quantum computing system, detailing the operation of a quantum-dot single-photon source at a specific temperature and repetition rate. This system utilizes an active demultiplexer and fibered delays to transform single photons into a simultaneous stream of 6 photons directed towards a 12-mode photonic chip. The chip, equipped with superconducting nanowire single-photon detectors, is controlled by a software stack to manipulate the unitary matrix $U$ via thermal phase shifters, ensuring recalibration for optimal performance. It houses a universal interferometer for versatile execution of $12\times 12$ unitary matrices. 
    (b) The N-photon coincidence rates over time, configured for the identity matrix, inclusive of monitoring system maintenance and assessing photon quality. The right-side figure monitors on-chip photon indistinguishability and single-photon purity through metrics like HOM (Hong-Ou-Mandel) visibility and normalized second-order correlation function~\citep{HOM}. 
    (c) The job execution process, delineating the submission, processing, and output retrieval for various computational tasks through Ascella.
    Reproduced with alternations under a Creative Commons Attribution 4.0 International License (\url{http://creativecommons.org/licenses/by/4.0/}) from~\citep{AscellaImageNature}.}
    \label{fig:Ascella}
\end{figure*}

Of particular note is \textit{Ascella}'s achievement in generating a distinct form of quantum entanglement involving three photons — heralded 3-photon entanglement~\citep{Ascella19,GBS,Ascella57,Univ12-mod13,Ascella59,Ascella60}. This accomplishment marks a pivotal step towards MBQC
~\citep{ORCAfltlrnt7} offering a promising trajectory for scaling up quantum computational systems~\citep{GBS0,quandela8}. 
Figure~\ref{fig:Ascella} portrays \textit{Ascella}'s architectural framework, performance characteristics, and stability.
For more detailed information about the structure and operation of the \textit{Ascella} quantum computer, including the processes of photon generation, manipulation, detection, and system control for implementing specific unitary matrices, readers are referred to~\citep{Ascella}.

\subsection{\textit{Perceval}}

Quandela presents \textit{Perceval}~\citep{Perceval}, an open-source software platform designed for the simulation and interfacing of discrete-variable photonic quantum computing systems. This innovative Python-based front-end enables the seamless composition of photonic circuits using fundamental building blocks, including photon sources, beam splitters, phase-shifters, and detectors. \textit{Perceval} offers a versatile selection of computational back-ends, each tailored to specific use-cases, utilizing state-of-the-art simulation techniques encompassing both weak and strong simulation~\citep{Perceval}. Through practical examples, \textit{Perceval}~\citep{Perceval} demonstrates its ability to reproduce photonic experiments and simulate a wide range of quantum algorithms, including Grover's~\citep{Grover1,Grover2,Grover3}, Shor's~\citep{Shor}, and QML. 
Targeted at experimentalists, theoreticians, and application designers, \textit{Perceval} stands as a valuable toolkit for modeling, designing, simulating, and optimizing discrete-variable photonic experiments and quantum computing applications, further bridging the gap between theory and practice in the field of quantum computing~\citep{ORCAfltlrnt23,Perceval,QuanAppl1,QuanAppl2,QuanAppl3,QuanAppl4,QuanAppl5,QuanQCom1,OneNine,QuanQCom3,Ascella,QuanQCom5,randomness,QuanQCom7,QuanQCom8,QuanQCom9,QuanQCom10,QuanQCom11,QuanQCom12,QuanQCom13,QuanQCom15}.

\subsection{Quandela Cloud}

Quandela offers a diverse array of functionalities designed to facilitate the utilization of its photonic quantum resources via Quandela cloud~\citep{QuanCLoud}. The platform presents a comprehensive display of the status and specifications of these quantum processors~\citep{OneNine}.

\input{mn}

\section{Conclusion}\label{Conclusion}

The burgeoning field of quantum computing, particularly within the realm of photonic architectures, stands as a beacon of technological promise. Its potential to reshape industries and redefine computational limits is undeniable. As we witness the strides made by industry leaders and the groundbreaking experiments unfolding before us, it becomes increasingly clear that we are on the precipice of a transformative era.

Photonic quantum computers are poised to dramatically enhance computational efficiency, outperforming classical systems in tasks such as material science simulations, cryptography, and optimization algorithms. Their practical applications could revolutionize secure communication protocols, improve precision in molecular simulations essential for drug discovery, and optimize complex logistical networks. Additionally, photonic quantum computers offer the ability to simulate intricate quantum phenomena with exceptional accuracy, advancing areas like quantum chemistry and materials science. Their capacity for complex optimization could significantly impact diverse sectors, including finance and energy management, underscoring their broad and transformative influence on computationally intensive industries.

This article captures a significant historical moment in the evolution of photonic quantum computers during the NISQ era, highlighting their potential to revolutionize the quantum computing landscape. By delving into the advancements made by key players in photonic quantum computing technology, exploring the current performance of their photonic quantum computers, their strategies for building large-scale fault-tolerant photonic quantum computers, and examining recent groundbreaking experiments, we underscore the remarkable potential of these technologies.

Through the fusion of photonics and quantum computing, we stand at the cusp of a new era in information processing, where the future of technology and science converges to redefine what is achievable in the quantum realm. This journey into the quantum realm signifies more than just a leap in computational power; it represents a paradigm shift, a redefinition of what is possible in the realms of technology and science. As we stand at the threshold of this new frontier, the potential for advancement and innovation knows no bounds. The fusion of photonics and quantum mechanics marks not only the dawn of a new era but also a testament to human ingenuity and our relentless pursuit of progress.



\input{Acknowledgements}

\input{references}
\end{document}

%% file: mn.tex
\section{The QuiX Quantum}

\subsection{The QuiX Quantum's photonic processors}

QuiX Quantum provides versatile multimode tunable interferometers in the form of photonic processors~\citep{QuixQuantum}. These photonic processors facilitate the arbitrary interference of light fields according to user-defined configurations. The fundamental structure of these processors is composed of optical waveguides, serving as conduits for light and defining distinct optical modes. Through controlled interactions between these waveguides, the intended optical operations are realized. The layout is meticulously designed to enable the implementation of a wide spectrum of linear optical transformations.

QuiX Quantum, founded in 2019, opts to utilize integrated photonics for its processor technology, leveraging the advantages of miniaturization. This approach ensures efficient scalability, maintains phase stability, and significantly reduces propagation loss, thanks to the utilization of a silicon nitride platform. The on-chip photonic processors by QuiX Quantum exhibit the capability to execute arbitrary linear operations with remarkable fidelity, rendering them suitable for a broad range of photonic applications.

In the current landscape, QuiX Quantum~\citep{QuixQuantum} stands out as a leading provider of integrated photonic processors~\citep{QuixRes1,QuixRes2,QuixRes3}. This recognition is due to their exceptional capability to manufacture processors of considerable size with minimal loss and exceptional fidelity. 
Presently, QuiX Quantum offers processors featuring a $20\times20$ mode configuration, and there are plans to further expand this capacity to $50\times50$ mode processors~\citep{Univ12-mod,quix12mod_InGaAs,Quix20mod}.

\subsection{The 12-mode quantum photonic processor}

\begin{figure*}
    \centering
    \includegraphics[width=\textwidth]{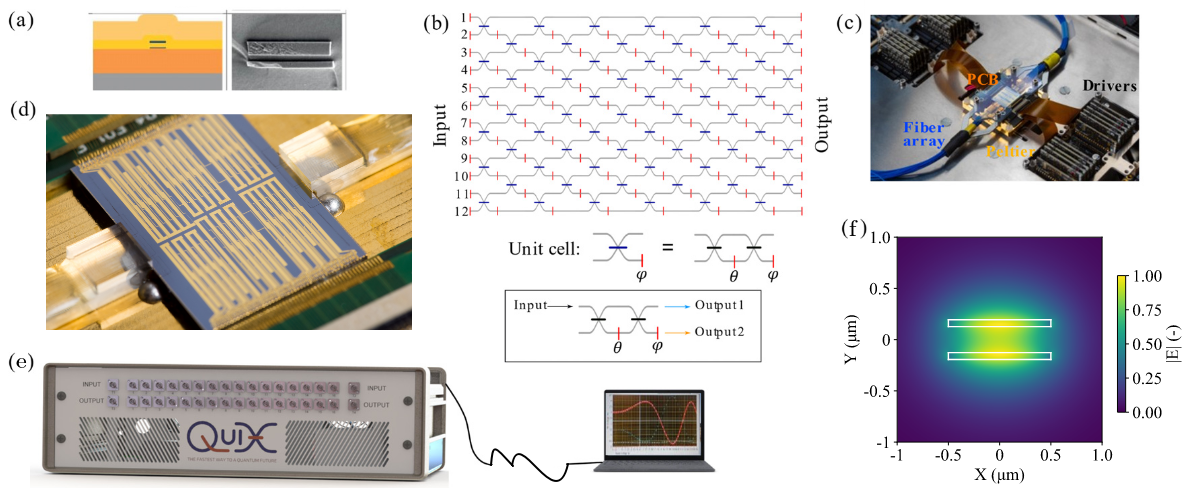}
    \caption{\textbf{An extensive view of the QuiX photonic processor.} In (a), a schematic and a scanning electron microscope (SEM) image illustrating the asymmetric double-stripe (ADS) cross-section utilized for the waveguides discussed in this research. (b) outline the functional design of the 12-mode photonic processor. Notably, the blue line signifies a tunable beam splitter (TBS) constructed as a Mach-Zehnder interferometer (MZI) incorporating two 50:50 directional couplers (depicted as black lines) along with a thermo-optic phase shifter (PS) indicated in red. During unit cell calibration, light is injected, for instance, through the top input, while monitoring both output channels. (c) offer a visual depiction of the photonic assembly of the 12-mode processor as situated within the control box. 
    (d) Depiction of the 12-mode photonic processor chip, measuring 16 mm $\times$ 22 mm~\citep{quix12mod_InGaAs}. 
    (e) presents a schematic representation of the QuiX control system, wherein the control box is remotely operated through a Python-based software interface. 
    Finally, (f) Representation of the simulated mode field profile for the supported TE mode at 940 nm, corresponding to the optimal waveguide design. The simulation employed a fully vectorial 2D eigenmode calculation conducted through Lumerical MODE solutions. The symmetric double-stripe SiN waveguide within the silica glass cladding is delineated by white lines. 
    Reprinted under the terms of the Creative Commons Attribution 4.0 license (\url{https://creativecommons.org/licenses/by/4.0/}) from~\citep{Univ12-mod}.
}
    \label{fig:Quix}
\end{figure*}

In~\citep{Univ12-mod}, Quix Quantum presented their successful demonstration of a universal quantum photonic processor—a highly versatile 12-mode fully tunable linear interferometer that enables comprehensive mode coupling while maintaining minimal loss. This processor is constructed using stoichiometric silicon nitride waveguides, and it comprises three key elements: an integrated silicon nitride photonic chip, associated peripheral equipment, and specialized control software to manage its operations.

The core of this photonic processor is a reconfigurable photonic integrated circuit that utilizes stoichiometric silicon nitride (Si$_3$N$_4$) waveguides with the TripleX technology~\citep{Quix20mod30}. These waveguides are engineered to achieve remarkably low propagation losses, as low as 0.1 dB/cm, and possess a minimum bending radius of 100 $\mu$m. The cross-sectional design of the waveguides utilized in the photonic processor is asymmetric double-stripe (ADS)~\citep{Quix20mod30}, as visually depicted in Figure~\ref{fig:Quix}(a). These waveguides are specifically engineered for single-mode operation at a wavelength of 1550 nm. The ADS waveguides facilitate efficient coupling to standard telecom fibers using spot-size converters, with the upper silicon nitride stripe being gradually tapered away through an adiabatic process~\citep{Quix20mod30}.

The reconfigurability of the photonic processor is achieved by harnessing the thermo-optic effect through resistive heating of 1 mm-long platinum phase shifters (PSs). Utilizing these thermo-optic PSs, a $\pi$ phase shift is achieved at a voltage of approximately $V_\pi \approx 10V$, corresponding to an electrical power consumption of roughly 385 mW per element~\citep{Univ12-mod}.

The functional architecture of the processor is presented in Figure~\ref{fig:Quix}(b). It consists of optical unit cells, which comprise a tunable 
TBS (depicted in blue) and a phase shifter (PS) (represented in red) on the bottom output mode. These unit cells are replicated 66 times within a square topology that supports 12 input/output modes (where 'modes' denote the zeroth order mode of each waveguide) and has a circuit depth of 12. The circuit depth is defined as the maximum number of unit cells encountered by an input mode in the direction of light propagation. Additionally, twenty-four extra PSs are strategically distributed across the inputs and outputs to facilitate sub-wavelength delay compensation and external phase adjustments. In total, the processor boasts 156 PSs. The TBSs are implemented through 
MZI, consisting of two $50:50$ directional couplers (as indicated by the black lines in Figure~\ref{fig:Quix}(b)) and an internal PS, denoted as $\theta$, followed by an external PS, marked as $\phi$, at the bottom output mode. Each unit cell effectively serves as a node within the large-scale interferometer, allowing for the interference of light~\citep{Univ12-mod1,Univ12-mod25,Univ12-mod26,Univ12-mod37}.

\subsection{The 20-mode quantum photonic processor}

\begin{figure}
    \centering
    \includegraphics[width=0.45\textwidth]{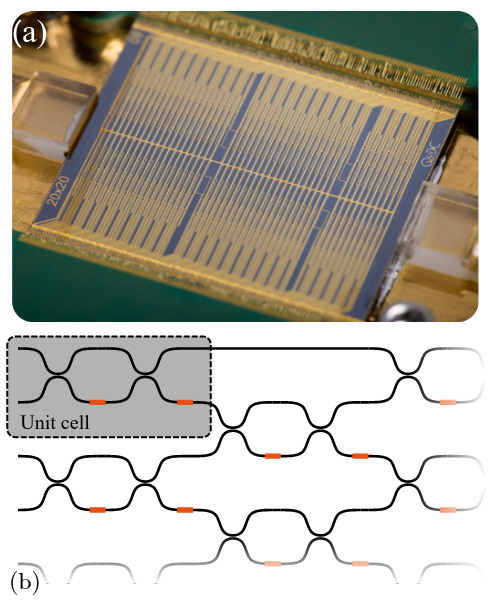}
    \caption{ \textbf{The 20-mod quantum photonic processor. }
    (a) A photographic representation of the 20-mode processor chip, measuring 22 mm $\times$ 30 mm. The chip undergoes optical packaging with an input/output fiber array and is wire-bonded to the control PCB, facilitating the targeted addressing of each tunable element. (b) The functional architecture of the processor, delineating the thermo-optic tunable elements (depicted in red) and the optimal configuration of waveguide paths (depicted in black). The mesh structure is established through the repetitive inclusion of unit cells, encompassing a tunable beam splitter (TBS) and an external phase shifter (PS), 
    iterated $\frac{1}{2}N(N - 1)$ times, where N equals 20, resulting in a cumulative total of 190 unit cells. 
    Reprinted under the terms of the 
Creative Commons Attribution (CC BY-NC-ND 4.0) International license (\url{ https://creativecommons.org/licenses/by-nc-nd/4.0/}) from~\citep{Quix20mod}.}
    \label{fig:QuiX20mod}
\end{figure}

In a recent publication~\citep{Quix20mod}, researchers from Quix Quantum, the University of Twente~\citep{MSA+}, and QuSoft~\citep{QuSoft} unveiled the most extensive universal quantum photonic processor known to date~\citep{Quix20mod}. This state-of-the-art processor empowers the execution of arbitrary unitary transformations on its 20 input/output modes, achieving remarkable amplitude fidelity rates of FHaar$ = 97.4\%$ for Haar-random matrices and FPerm$ = 99.5\%$ for permutation matrices~\citep{Quix20mod}. Notably, the optical losses across all operational modes maintain an average of 2.9 dB, while the processor exhibits high-visibility quantum interference at VHOM$ = 98\%$. It is pertinent to highlight that the implementation of this processor relies on Si3N4 waveguides and is effectively temperature-regulated by means of a Peltier element~\citep{Quix20mod}.

The Quix Quantum processor~\citep{Quix20mod} is ingeniously engineered employing Si$_3$N$_4$ waveguides, underpinned by TripleX technology~\citep{Quix20mod30}. To maintain consistent temperature levels, it is bolstered by a water-cooled Peltier element. These waveguides feature an asymmetric double-stripe cross-sectional design and possess a minimal bending radius of 100 $\mu$m. Impressively, the device records an average loss of 2.9 dB across all modes, and it accommodates arbitrary linear optical transformations, making it compatible with diverse linear optical models of quantum computation~\citep{GBS0}. Furthermore, the reconfigurability of the processor was subjected to rigorous testing, encompassing more than 1000 unitary transformations. The results of this testing underscore the remarkable degree of control exhibited by the linear optical interferometer and affirm its universality~\citep{Univ12-mod26}. The processor's performance was further substantiated through the execution of 190 quantum interference experiments, conclusively affirming its competence in preserving the inherent properties of quantum light at every constituent component~\citep{Quix20mod}.

The 20-mode quantum photonic processor comprises three pivotal constituents: the Si$_3$N$_4$ photonic chip, the peripheral system, encompassing control electronics, and a dedicated control software suite~\citep{Univ12-mod}. The Si3N4 photonic chip (as depicted in Figure~\ref{fig:QuiX20mod}~(a) features a comprehensive assembly of 380 thermo-optic tunable elements, structured within a universal square interferometer (as illustrated in Figure~\ref{fig:QuiX20mod}~(b). Each unit cell within this configuration consists of a TBS 
succeeded by a PS~\citep{Univ12-mod26}. Impressively, the chip manifests minimal propagation losses, registering as low as 0.07 dB/cm at a wavelength of 1562 nm, thanks to an upgraded annealing process compared to their preceding chip~\citep{Univ12-mod}. The peripheral system encompasses the control electronics, and an active cooling module complements the setup. The thermo-optic tunable elements can be rapidly switched at a kHz rate~\citep{Quix20mod30}, establishing a benchmark for the transition speed between various processor configurations. To uphold precise temperature control over the photonic chip, an active cooling mechanism is employed, entailing a Peltier element connected to a water cooling module, delivering a maximum heat reduction rate of 200 W~\citep{Quix20mod}.

Table~\ref{tab:12compa20} presents a comparison of the Quix Quantum two photonic quantum processor modes: the 12-mode~\citep{Univ12-mod} and 20-mode~\citep{Quix20mod}. It highlights key performance metrics, including the number of thermo-optic phase shifters (represented as``PSs"), Insertion Loss (IL) in decibels (dB), Coupling Loss (CL) in decibels per facet (dB/facet), and Propagation Loss (PL) in decibels per centimeter (dB/cm) for both processor configurations~\citep{Univ12-mod,Quix20mod}.

The Quix Quantum's potential signifies a significant stride towards the realization of low-loss, fully reconfigurable linear optical interferometers, known as, Quantum Photonic Processors (QPPs)~\citep{Univ12-mod,Univ12-mod8,Univ12-mod1,Univ12-mod7,Univ12-mod4}, a pivotal development in the domain of quantum computing and information processing.
The Quix Quantum preferred its quantum processor~\citep{Quix20mod} over 
other impressive works of large-scale photonic integrated circuits~\citep{Univ12-mod7,Univ12-mod12,Univ12-mod3,Quix20mod42,Quix20mod43,Univ12-mod22,Quix20mod45,Quix20mod46,Quix20mod47,Quix20mod48,Quix20mod49,Quix20mod50,Quix20mod51,Quix20mod52,Univ12-mod2,Univ12-mod10,Univ12-mod11,Quix20mod56,PsiQ20}, as they are either non-universal or have unrecorded performance in terms of losses~\citep{Quix20mod}.

\begin{table}
\caption{A comparison between the characteristics of 12-mode~\citep{Univ12-mod} and 20-mode~\citep{Quix20mod} quantum photonic processors (QPPs). The table provides information on the number of thermo-optic phase shifters (PSs), Insertion Loss (IL), Coupling Loss (CL), and Propagation Loss (PL) for both processor modes.}
\label{tab:12compa20}
\resizebox{0.5\textwidth}{!}{
\centering
\begin{tabular}{lcccc}
\hline \hline
\textbf{QPPs} & \textbf{PSs} & \textbf{CL (dB/facet)} & \textbf{IL (dB)}& \textbf{PL (dB/cm)} \\
\hline 
The 12-mode  & 132  & 2.1 & 5.0& 0.1 \\
The 20-mode  & 380  & 0.9 & 2.9& 0.07 \\
\hline \hline
\end{tabular}
}
\end{table}

\section{TundraSystems Global}

\subsection{Overview}

TundraSystems Global~\citep{TundraSystems}, is a photonic quantum computing company established in 2014. 
The overarching goal of TundraSystems Global is to pioneer and deliver cutting-edge quantum technology solutions. Their initial developmental phase focuses on creating the Tundra Quantum Photonics Technology library. This library is an integral component of Tundra System's strategic vision, aiming to realize a comprehensive quantum photonics microprocessor (QPM) known as the TundraProcessor~\citep{TundraSysInsights}. 
The establishment of this library is intended to support the development of a broader ecosystem of photonic integrated circuits, facilitating the construction of complete High-Performance Computing (HPC) systems centered around the TundraProcessor~\citep{TundraSysInsights,TundraSysQPP}.

\subsection{TundraSystems' quantum computing ecosystem}

TundraSystems' product portfolio comprises a comprehensive quantum computing stack encompassing both hardware and software components~\citep{TundraSysQSoftw}. Notably, it incorporates an integrated quantum error correction system based on deep learning techniques~\citep{TundraSysError}. At its core, the quantum systems offered feature a 64-qubit quantum processor~\citep{TundraSysInsights}. These systems are designed to operate as HPC units and are intended to address a wide range of HPC and quantum computing requirements~\citep{TundraSysInsights,TundraSysError,TundraSysQPP,TundraSysQSoftw}.

\subsection{The versatile TundraSystem HPC units}

The TundraSystem HPC units represent complete quantum computers, incorporating a quantum processor founded on Silicon photonics technology and accompanying hardware peripherals~\citep{TundraSysQPP}. The software layers encompass the Tundra quantum instruction Set architecture (QISA), the Tundra quantum operating system (TundraQOS), and the Tundra cross compiler. These software elements empower users to seamlessly migrate or adapt their software algorithms from diverse platforms onto the Tundra Platform~\citep{TundraSystems,TundraSysQSoftw}.

An additional distinctive feature of each TundraSystems' HPC unit is the inclusion of deep learning-based quantum error correction (TundraQECDL)~\citep{TundraSysError}. This solution not only serves as the organization's standalone quantum error correction (QEC) system but is also a versatile and adaptable error correction tool applicable to any quantum technology solution, whether it pertains to computing or communication, and necessitates error correction~\citep{TundraSysError}.

\section{TuringQ}

\begin{figure*}
    \centering
    \includegraphics[width=0.95\textwidth]{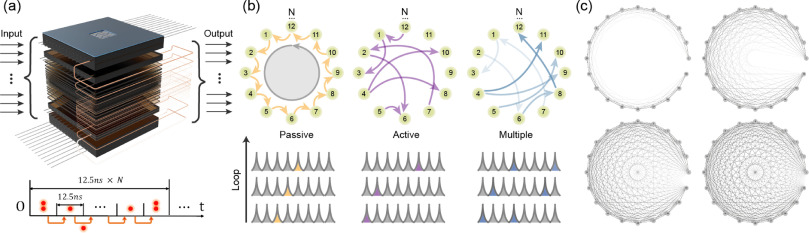}   \\
    \vspace{0.5cm}
    \includegraphics[width=0.9\textwidth]{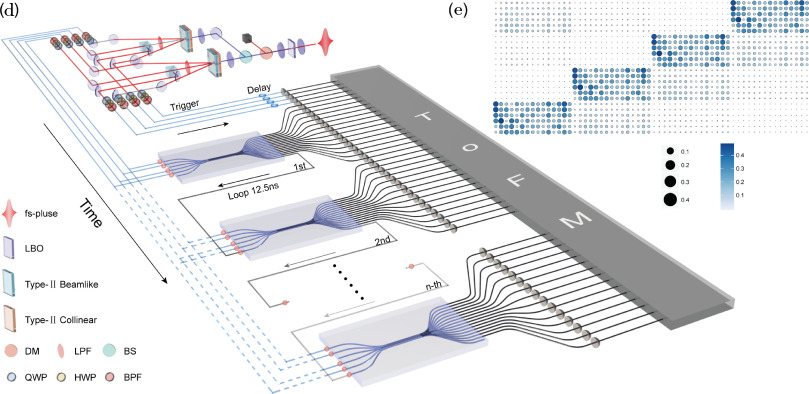}
    \caption{
    \textbf{The membraneboson sampling scheme} (a) Each discrete constituent unit of the boson sampling framework is characterized by the matrix $\Lambda$. Leveraging a memristor-like effect, the scale of the boson sampling problem can be expanded infinitely. The timeline delineates the sequential progression of photons through various temporal layers. (b) Potential modifications to the loop architecture are presented, encompassing three distinct loop-based structures: passive loop, active loop, and multiple loops. (c) The graph structure associated with membosonsampling is elucidated, where each node represents an individual layer, and the edges denote potential transitions between different layers. The complexity of the structure is proportionate to the number of conceivable transition edges.
    A schematic of the membosonsampling machine, known as``Zhiyuan": (d) The generation of multi-photon Fock states involves successive pumping of $\beta$-barium borate (BBO) crystals with a frequency-doubled femtosecond laser pulse at 390nm. Photons heralded from down-converted pairs are collected via fiber couplers and directed to the photonic chip. Each time interval serves as resources for both photons and circuit modes, enabling scalability. The loop structure facilitates quantum interference across different temporal layers. The output in various layers is detected by an extensive array of avalanche photodiodes (APDs). Electronic signals are recorded by the time-of-flight module (TOFM), which captures time information in the large Hilbert dimensions of the state space simultaneously. Trigger channel time information heralds corresponding photon coincidence events. (e) Experimental characterization illustrates the scattering matrix of a 4-layer structure, demonstrating nonzero survival probabilities even after several cascaded layers, thereby enabling inter-layer quantum interference. 
    Reproduced under the Creative Commons Attribution NonCommercial License 4.0 (\url{https://creativecommons.org/licenses/by-nc-nd/4.0/}) from~\citep{Xian-Jin5}.}
    \label{fig:membranebosonsampling}
\end{figure*}

\begin{figure*}
    \centering
    \includegraphics[width=0.9\textwidth]{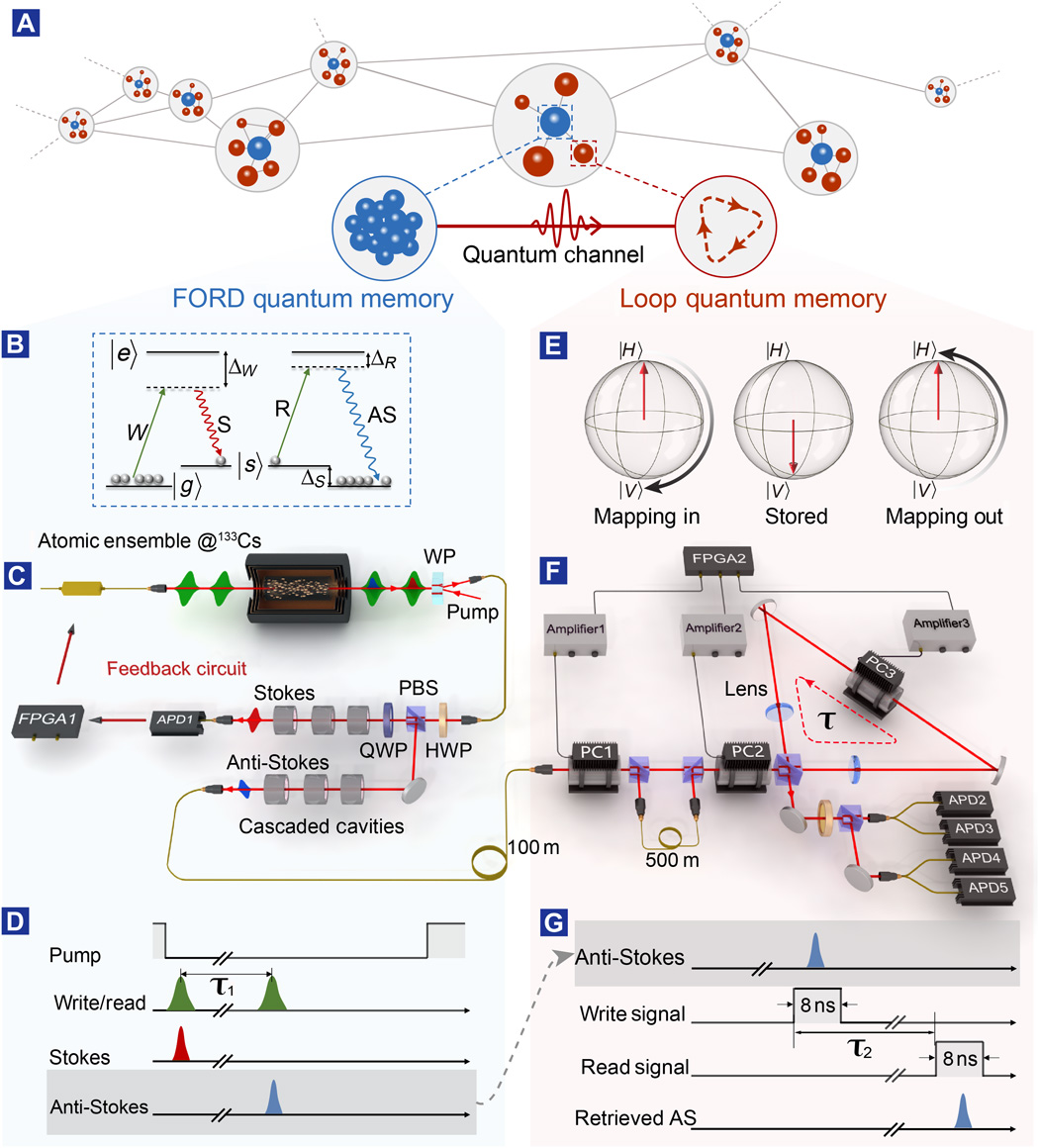}
    \caption{\textbf{Schematic and Experimental Configuration of a Hybrid Quantum Memory-Enabled Network. }
(A) The quantum network comprises two distinct functional nodes interconnected for hybrid quantum memory. 
(B) The write and read processes of the FORD quantum memory involve a three-level $\Lambda$-type atomic configuration in cesium atoms, where $\ket{g}$ and $\ket{s}$ denote hyperfine ground states ($\Delta_s = 9.2$ GHz), and $\ket{e}$ represents the excited state. Dashed lines indicate broad virtual energy levels induced by write and read pulses. 
(C) The FORD quantum memory setup includes a Wollaston prism (WP), polarization beam splitter (PBS), quarter-wave plate (QWP), and half-wave plate (HWP). 
(D) Time sequences of the FORD quantum memory. (E) Polarization switching during the mapping in-and-out processes is illustrated in Bloch spheres. 
(F) The Loop quantum memory setup involves a Pockels cell controlled by write and read electrical signals from a field-programmable gate array (FPGA) module. A 500-m-long fiber acts as an additional switching path against photon loss, with four avalanche photodiodes (APDs) for photon detection at small time intervals. PC denotes Pockels cells. 
(G) Time sequences of the Loop memory, where the time interval $\tau_2$ between write and read signals can be any positive integral multiples of one cycle period $\tau$. 
Reproduced under the Creative Commons Attribution NonCommercial License 4.0 (\url{ https://creativecommons.org/licenses/by-nc/4.0/}) from Ref.~\citep{Xian-Jin3}.}
    \label{fig:quantumnetwork}
\end{figure*}

\begin{figure*}
    \centering
    \includegraphics[width=0.9\textwidth]{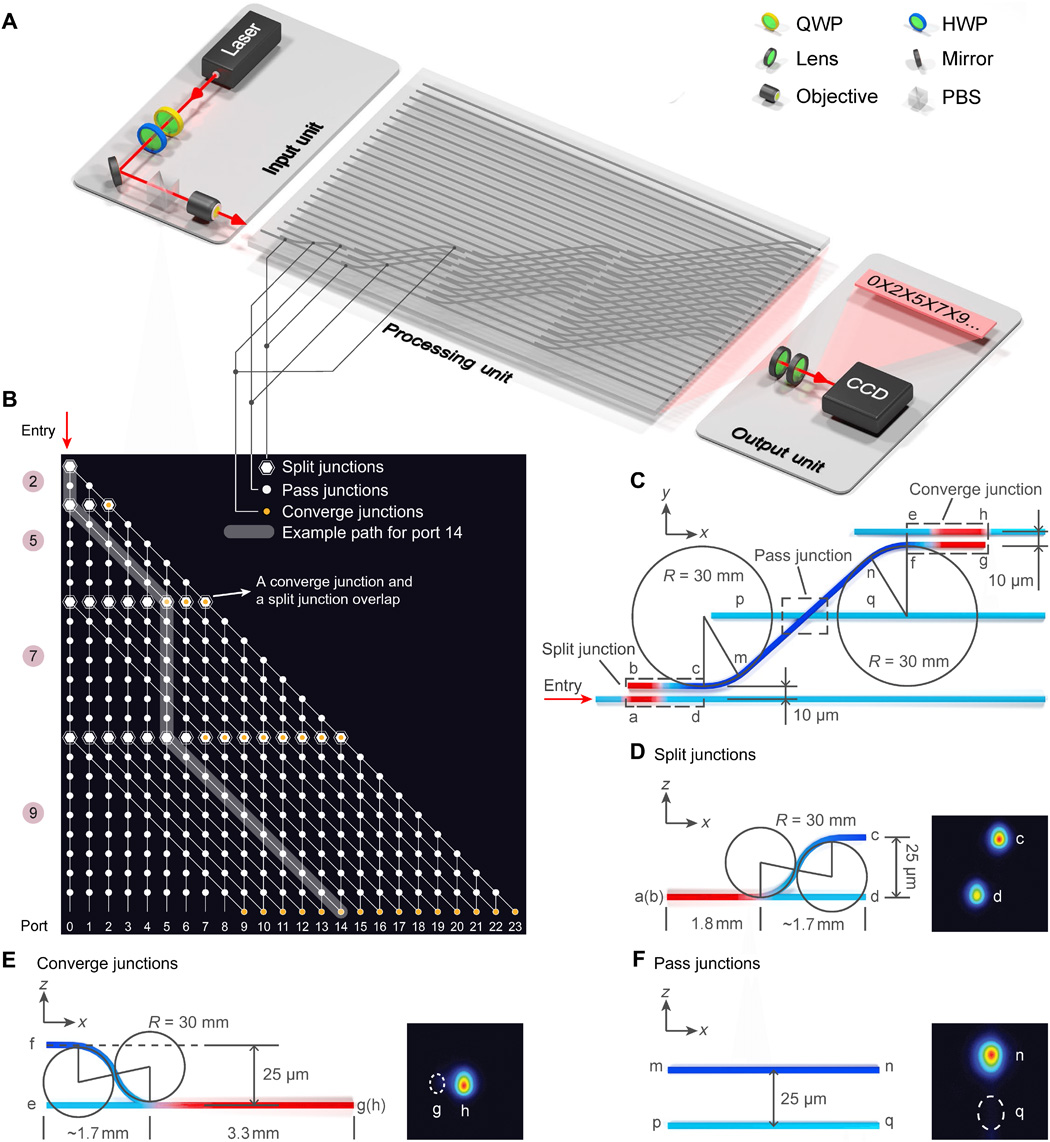}
\caption{\textbf{Design and Setup Overview. }
(A) The input unit employs a power-adjustable and horizontally polarized optical source achieved through a quarter-wave plate (QWP), half-wave plate (HWP), and polarization beam splitter (PBS). Photons at 810 nm are prepared and coupled in the processing unit, generating all possible subset sums. The resulting outcomes at the output ports are captured by the charge-coupled device (CCD) for validation.
(B) The abstract network for the specific set $\{2, 5, 7, 9\}$ includes split junctions (hexagonal nodes), pass junctions (circular white nodes), and converge junctions (circular yellow nodes). Split junctions divide photon streams into vertical and diagonal segments, while pass junctions maintain original trajectories, and converge junctions transfer photons from diagonal to vertical paths. Despite visual overlap, the physical separation of circular yellow nodes from hexagonal nodes is evident in (A). Photons moving diagonally between split junctions represent elements in the summation, with values corresponding to the number of junctions between successive rows of split junctions. Subset sums align with the spatial positions of output signals.
(C) The $x-y$ view of the top-left corner of the waveguide network and abstract network features three junction types, with $x-z$ views in (D) to (F). The split junction employs a modified three-dimensional beam splitter with a 10 $\mu$m coupling distance, 1.8 mm coupling length, and 25 $\mu$m vertical decoupling distance. Unbalanced output compensates for bending losses induced by subsequent arcs $\widehat{cm}$ and $\widehat{nf}$ in (C). The converge junction mirrors the split junction but with a 3.3 mm coupling length. Negligible residual in port g and a 25 $\mu$m vertical decoupling distance ensure an excellent pass junction with an extinction ratio of approximately 24 dB, visible in the intensity distribution in (F). 
Reproduced under the   Creative Commons Attribution NonCommercial License 4.0 (\url{https://creativecommons.org/licenses/by-nc/4.0/}) from Ref.~\citep{Xian-Jin4}.}
    \label{fig:Design}
\end{figure*}

\subsection{Overview}

TuringQ~\citep{TuringQ}, was established in 2021, positioning itself as the first Chinese company dedicated to optical quantum computer chip development. The core of their work involves the creation of extensive photonic circuits using lithium niobate on insulator (LNOI) photonic chips and femtosecond laser direct writing technology~\citep{Zewail1,femtosecondWriting,femtolaser}. 
Despite its relative youth, TuringQ has introduced a range of innovative products to the market. Notable offerings include the TuringQ Gen 1, a research-grade optical quantum computer optimized for commercial applications, a 3D optical quantum chip, and a high-speed programmable optical quantum chip. Among their portfolio is FeynmanPAQS, a commercially viable, self-developed optical quantum computing simulation software~\citep{TuringQSoft}.

\subsection{TurinQ optical quantum computing systems}

The TurinQ quantum computer is a commercial optical quantum computing platform~\citep{TuringQ}. This quantum computer's configuration comprises several key components, including: 

\textit{Quantum Light Source }
At its core, this system incorporates a Sagnac interferometer-based quantum light source, utilizing a continuous, narrow linewidth semiconductor laser emitting at a $405$nm wavelength. This laser, combined with meticulous temperature control maintained at a  precision level of 0.001K and a stable elastic vibration isolation system, produces a high-quality, ultra-bright quantum light source, boasting an output brightness exceeding $500,000$ pairs per second~\citep{TuringQSoft}.

\textit{Detection Systems }
This system boasts high temporal resolution with response times measuring less than 2 nanoseconds per billion parts (bps), enabling precise investigations of transient phenomena. Additionally, its high-speed readout platform, operating at speeds of up to $5$MHz, incorporates high-precision gate control with low latency and an exceptional $10$ picoseconds set-up accuracy, which proves invaluable for complex experimental integrations~\citep{TuringQSoft}.

\textit{Temperature Control System }
With extraordinary temperature control accuracy, reaching levels as low as $0.001$K, ensuring precise and reliable thermal management for quantum operations~\citep{TuringQSoft}.

\textit{Damping System }
The damping system in this quantum computer employs a high damping solution characterized by a loss factor ranging from approximately $0.5$ to $0.8$ within the frequency range of $10-100$~Hz at room temperature~\citep{Xian-Jin3}. This performance is claimed to surpass that of other commercial damping materials by two to five times, emphasizing the system's exceptional efficiency in vibration control~\citep{TuringQSoft}.

\textit{Interface Control System }
The interface control system, encompasses various critical components, offers users a versatile platform for tailored secondary development, highlighting the flexibility and adaptability of the TurinQ quantum computer to meet diverse research and application needs~\citep{Qcrypto2002,errorcorr2013,NISQ23}.

\subsection{Quantum computational advantage with membosonsampling}

Boson sampling has emerged as a leading candidate for demonstrating quantum advantage~\citep{GBS0}, particularly with applications proposed based on its computational problem~\citep{NISQ23}. Numerous protocols have been proposed for real-life applications, leveraging the core concept of boson sampling~\citep{XanSoftw26,XanSoftw28,memboson12,memboson13}. Large-scale boson sampling experiments, ranging from seminal prototypes~\citep{memboson15,memboson16,memboson17,Univ12-mod1,memboson19,memboson20,Ascella59,memboson22,memboson23,memboson24,memboson25,Ascella62,Univ12-mod10,Univ12-mod11,memboson29,memboson30,Ascella63,memboson32}, to variants using different input states like scattershot boson sampling~\citep{memboson33,Ascella57} and GBS~\citep{Jiuzhang,Univ12-mod13}, further underscore the significance of this approach.

\textit{Quantum advantage with membosonsampling} In~\citep{Xian-Jin5}, Jun Gao \textit{et al.} puts forth and experimentally demonstrates quantum advantage utilizing the membosonsampling machine named ``\textit{Zhiyuan}". The experimental verification is conducted on a self-looped photonic chip inspired by the memristor concept. Through the establishment of quantum interference between different temporal layers, the researchers enable the scaling up of the boson sampling problem to dimensions surpassing the tractable capacity of classical supercomputers~\citep{memboson45,memboson46}. The achieved multi-photon registrations reach up to 56-fold in a system with $750,000$ modes, encompassing an exponentially large computational Hilbert space extending to $10^{254}$~\citep{Xian-Jin5}. These outcomes signify an integrated and cost-efficient approach, representing a substantial leap into the ``quantum computational advantage" regime~\citep{NISQ23,nisqQC10} within a photonic system. Moreover, the presented platform offers scalability and controllability, serving as a promising foundation for QIP~\citep{Nilson,qc}. 
The membraneboson sampling scheme as well as a schematic of the membosonsampling machine,  \textit{Zhiyuan} are illustrated in Figure~\ref{fig:membranebosonsampling}. 
Beyond the realm of quantum computing, TuringQ may explore numerous perspectives on fully leveraging this straightforward method of constructing large-scale quantum systems within the broader context of quantum sciences and technologies. Detailed information about the procedure on extracting multi-photon events in membosonsampling is provided in~\citep{Xian-Jin5}.

\subsection{Experimental quantum fast hitting on hexagonal graphs}

\textit{Two-dimensional quantum walk on a photonic chip} In ~\citep{Xian-Jin1}, H. Tang \textit{et al.} have exemplified the realization of a two-dimensional continuous-time quantum walk, harnessing the external geometry of photonic waveguide arrays as opposed to the intrinsic degrees of freedom of photons~\citep{Xian-Jin1}. By employing femtosecond laser direct writing techniques~\citep{Zewail1,femtosecondWriting,femtolaser}, they fabricated a substantial three-dimensional structure that configures a two-dimensional lattice featuring up to $49 \times 49 $ nodes on a photonic chip~\citep{Xian-Jin1}. They subsequently illustrated the execution of spatial two-dimensional quantum walks using heralded single photons and single-photon-level imaging~\citep{Xian-Jin1}. 

\textit{Experimental quantum fast hitting on hexagonal graphs} Furthermore, H. Tang \textit{et al.}~\citep{Xian-Jin2}  conducted experimental demonstrations of quantum fast hitting by executing two-dimensional quantum walks on graphs comprising as many as 160 nodes and extending to a depth of eight layers. Remarkably, they established a direct correlation between the optimal hitting time and the network depth, thereby paving the way for a scalable approach towards achieving quantum acceleration in the context of classically intractable complex problems~\citep{Xian-Jin2,memboson48}.

\subsection{Quantum network with hybrid quantum memory}

\textit{A hybrid quantum memory} The development of quantum information technologies necessitates quantum memory systems proficient in the on-demand storage and retrieval of flying photons. Nevertheless, the requisite devices for establishing long-distance quantum links differ from those envisaged for localized processing. 
Numerous investigations have been undertaken in the realm of quantum memory theory and its associated physical realizations, encompassing electromagnetically induced transparency (EIT)~\citep{Qmemory18,Qmemory19,Qmemory20}, optical delay lines and cavities ~\citep{Qmemory14,Qmemory15,Qmemory16,Qmemory17,Algorithms}, the Duan-Lukin-Cirac-Zoller (DLCZ) protocol~\citep{Qmemory8,Qmemory21,Qmemory22}, off-resonant Faraday interaction~\citep{Qmemory25}, photon-echo quantum memory~\citep{Qmemory23,Qmemory24}, Raman memory~\citep{Qmemory26,Qmemory27}, and Autler-Townes splitting (ATS) memory~\citep{Qmemory28}. Significantly, endeavors have been directed toward expanding bandwidth from the kilohertz and megahertz range to the gigahertz range, as well as extending the operational temperature from near absolute zero to room temperature~\citep{Qmemory18,Qmemory19,Qmemory20,Qmemory21,Qmemory22,Qmemory23,Qmemory25,Qmemory26,Qmemory27}. Persistent endeavors continue to enhance the performance and practicality~\citep{Qmemory9,Qmemory10,Qmemory12,Qmemory13,Qmemory15,Qmemory16,Qmemory17,Qmemory31,Qmemory32,Qmemory33,Qmemory35}.

In~\citep{Xian-Jin3}, Pang \textit{et al.} propose and validate a versatile quantum network with hybrid quantum memory. The network's core components include an atomic ensemble-based memory for generating and storing quantum states, and an all-optical memory for mapping incoming photons bidirectionally at room temperature with high bandwidth~\citep{Xian-Jin3}, as shown in Figure~\ref{fig:quantumnetwork}. These complementary quantum memories operate compatibly at room temperature, facilitating the generation, storage, and conversion of atomic excitations into broadband photons. Despite the current microsecond-level lifetime limitation~\citep{Xian-Jin3,Qmemory43}, this hybrid quantum network is already practical for local quantum computers and simulators~\citep{qc,Qmemory5,Qmemory6,Qmemory7}.

\subsection{Photonic computer solving the subset sum problem}

Despite the formidable challenges, certain researchers endeavor to address NP-complete problems~\citep{SubsetSum1} within polynomial time and resource constraints. Demonstrations of memcomputing machines~\citep{SubsetSum6,SubsetSum7} as powerful as non-deterministic Turing machines (NTMs) have been achieved, albeit with acknowledgment of impracticalities in real-world scenarios due to inevitable noise~\citep{SubsetSum8}. Proposed designs for NTMs, incorporating magical oracles to explore all computation paths simultaneously~\citep{SubsetSum1,SubsetSum9,SubsetSum10}, come at the expense of space or material but offer an alternative to reduce time consumption~\citep{SubsetSum11,SubsetSum12}. 
In addition to NTM proposals, parallel exploration has been employed in various technologies, such as electronic supercomputers integrating an increasing number of processors~\citep{SubsetSum13} and molecular-based computation utilizing large quantities of DNAs or motor molecules~\citep{SubsetSum14,SubsetSum15,SubsetSum16,SubsetSum17,SubsetSum18} optimized algorithms have been applied to specific instances~\citep{SubsetSum19,SubsetSum20,SubsetSum21}.

Photons have played a significant role in proof-of-principle demonstrations of supercomputing~\citep{NISQ23,SubsetSum27}, even without quantum speed-up. These demonstrations include addressing NP problems like prime factorization~\citep{SubsetSum8} and NP-complete  Hamiltonian path problem~\citep{SubsetSum30,SubsetSum31,SubsetSum32}, problems like the traveling salesman problem~\citep{SubsetSum29}, and dominating set problem~\citep{SubsetSum33}. The photonic regime has also witnessed successful applications in solving \#P-complete problems like boson sampling~\citep{Univ12-mod11,Univ12-mod10,memboson16,Quix20mod51,memboson17,Univ12-mod1}, as well as other computational functions ~\citep{SubsetSum40} and algorithms~\citep{SubsetSum41,Xian-Jin2}, indicating the potential of photons as promising candidates for solving the subset sum problem (SSP).

\textit{Solving the subset sum problem} The subset sum problem (SSP), a challenging NP-complete problem for classical computers~\citep{SubsetSum1,SubsetSum2,SubsetSum3}, can be efficiently addressed using photons due to their high speed, robustness, and low energy detectability. In~\citep{Xian-Jin4}, Xu \textit{et al.} introduce a chip-based photonic computer that maps SSP onto a three-dimensional waveguide network using femtosecond laser direct writing~\citep{Zewail1,femtosecondWriting,femtolaser}. This approach, particularly effective for successive primes, outperforms supercomputers in terms of time consumption.  A schematic representation of the design and setup is illustrated in Figure~\ref{fig:Design}. These findings highlight light's ability to tackle computations beyond classical computer capabilities, positioning SSP as a benchmark for assessing the competition between photonic and classical computers toward achieving``photonic supremacy." For a comprehensive understanding of TuringQ's research, technology, programmable photonic chip, chip simulation software, and quantum cloud SoftQubit, interested readers are directed to~\citep{TuringQ,Xian-Jin1,Xian-Jin2,Xian-Jin3,Xian-Jin4,Xian-Jin5}.

\section{Xanadu Quantum Technologies}\label{xanadu}

\subsection{Overview}

Xanadu~\citep{Xanadu} has unveiled a groundbreaking achievement with the introduction of \textit{Borealis}~\citep{Borealis}. \textit{Borealis} emerges as a fully programmable photonic quantum computing system that showcases quantum computational advantage~\citep{Borealis}. Since its founding in 2016, Xanadu's commitment to advancing quantum research is evidenced by their open-source library, \textit{PennyLane}, which seamlessly integrates QML, 
computing, and chemistry, and can interface with various quantum processors, including those by IBM~\citep{ibm} and Rigetti~\citep{rigetti}.

\subsection{\textit{Borealis}}

\begin{figure*}
    \centering
        \includegraphics[width=\textwidth]{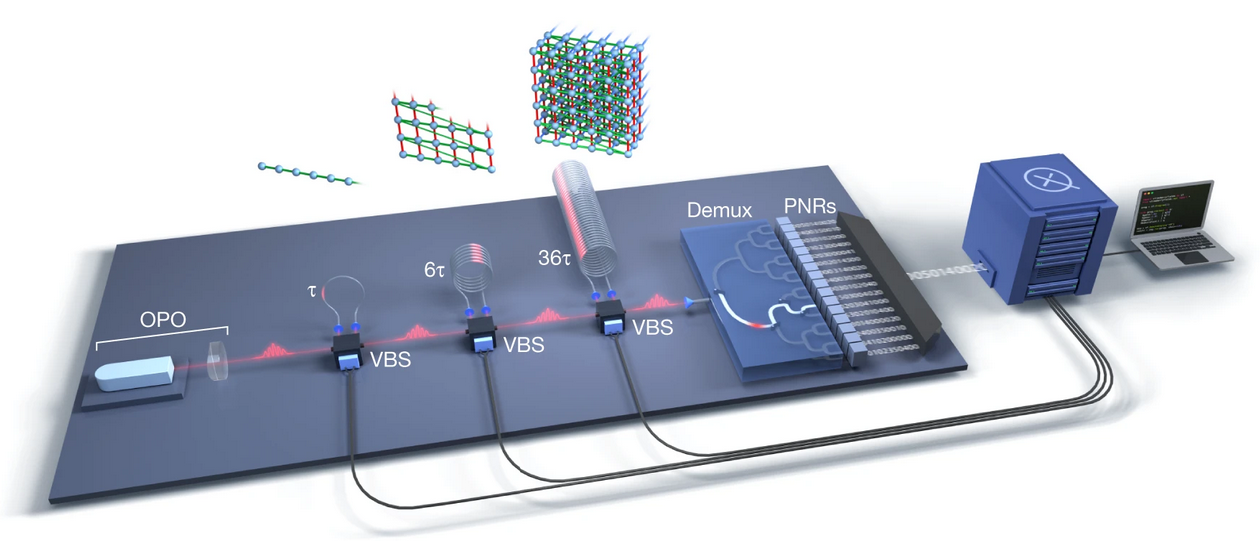}
\caption{\textbf{Attaining quantum computational advantages with programmable photonic quantum processor, \textit{Borealis}.} 
A sequence of single-mode squeezed states is generated by a pulsed OPO, forming periodic pulse trains. These states are directed into a series of three dynamically programmable loop-based interferometers. Each loop comprises a VBS with a programmable phase shifter and an optical fiber delay line. The output of the interferometer is directed to a 1-to-16 binary switch tree (demux) that partially separates the output before being read by PNRs. The outcome is a detected sequence of 216 photon numbers, obtained in approximately 36 $\mu$s, constituting one sample. The implemented fiber delays, beamsplitters, and phase shifters enable gates between adjacent and distant modes, allowing for extensive connectivity in the quantum circuit. Above each loop stage, a lattice representation illustrates the gradual synthesis of multipartite entangled Gaussian states. The first stage ($\tau$) introduces two-mode programmable gates (green edges)
, connecting nearest-neighbor modes in one dimension. The second and third stages ($6\tau$ and $36\tau$ respectively) establish connections between modes separated by six and 36 time bins in the second and third dimensions, (respectively the red and the blue edges)
. Each device run necessitates the specification of 1,296 real parameters, corresponding to the settings for all VBS units. Reproduced 
under a Creative Commons Attribution 4.0 International License (\url{http://creativecommons.org/licenses/by/4.0/})  from Ref.~\citep{Borealis}.}
\label{fig:Borealis}
\end{figure*}

A significant advancement is showcased  wherein quantum computational advantage is demonstrated by \textit{Borealis}~\citep{Borealis}, a photonic quantum processor renowned for its dynamic programmability across all implemented gates. The research employs 
GBS~\citep{GBS} employing 216 squeezed modes entangled through a three-dimensional connectivity configuration~\citep{Borealis5}. This is accomplished using a time-multiplexed and photon-number-resolving architecture. The key finding reveals that \textit{Borealis} accomplishes this task within a mere 36 microseconds, while the best available algorithms and supercomputers would purportedly require more than 9,000 years to achieve the same result~\citep{Borealis}.  This performance improvement is exceedingly substantial, exceeding the advantage demonstrated by earlier photonic machines by over 50 million times~\citep{Borealis}. 
An illustration of this photonic quantum processor is shown in Figure~\ref{fig:Borealis}.

The research notably represents an extensive GBS experiment, recording events involving up to 219 photons with a mean photon number of 125. This marked achievement is a significant milestone toward the realization of a practical quantum computer. It validates crucial technological aspects of photonics as a viable platform for advancing towards this goal. The accomplishment of achieving quantum computational advantage through Xanadu's photonic quantum computer, \textit{Borealis}, is examined in~\citep{Borealis}. Furthermore, these devices were susceptible to spoofing~\citep{Borealis3}, a phenomenon wherein classical heuristics generated samples that appeared closer to the desired distribution than those obtained from the quantum hardware itself.

\subsection{\textit{Borealis}'s optical circuit design}

The optical interferometer system takes its input from a single optical parametric oscillator (OPO) that emits pulsed single-mode squeezed states at a rate of 6 MHz. These states are subsequently directed into three concatenated, programmable, loop-based interferometers. Each loop comprises a Variable Beamsplitter (VBS), featuring a programmable phase shifter, along with an optical fiber delay line that functions as a buffer memory for light. The loops enable the interference of modes that are temporally adjacent, with a temporal separation of $\tau = (6 \text{ MHz})^{-1}$, as well as modes separated by either six or 36 time bins ($6 \tau$ or $36 \tau$) in the first, second, and third loop, respectively.

The optical delays serve as an effective means to facilitate both short- and long-range couplings between modes. For this experiment, a high-dimensional Gaussian state is generated, and its visualization is provided above the three loops in Figure~\ref{fig:Borealis}, represented in a three-dimensional lattice configuration. Assuming a lattice of size $a = 6$, where $a$ signifies the number of modes between two interacting pulses in the second loop, it is possible to construct a cubic lattice by introducing $M = a^3 = 216$ squeezed-light pulses into the interferometer.

\subsection{\textit{PennyLane}}

\textit{PennyLane}~\citep{pennylane,XanQML22} is a Python 3 software framework designed for differentiable programming in the realm of quantum computing. It presents a unified architecture that caters to near-term quantum computing devices, encompassing both qubit and continuous-variable paradigms. The distinctive feature of \textit{PennyLane} lies in its capacity to compute gradients of variational quantum circuits in a manner congruent with classical techniques such as backpropagation. Consequently, \textit{PennyLane} extends the reach of automatic differentiation algorithms commonly utilized in optimization and machine learning to encompass quantum and hybrid computations. A versatile plugin system ensures the framework's compatibility with a wide range of gate-based quantum simulators and hardware platforms. \textit{PennyLane} offers plugins for hardware providers, including the Xanadu Cloud, Amazon Braket~\citep{AWS}, and IBM Quantum~\citep{ibm}, enabling the execution of \textit{PennyLane} optimizations on publicly accessible quantum devices. On the classical computing front, \textit{PennyLane} interfaces seamlessly with accelerated machine learning libraries, including TensorFlow, PyTorch, JAX, and Autograd. It finds applications in the optimization of VQES (variational quantum eigensolvers), QAO (quantum approximate optimization), QML models, and a myriad of other quantum-related applications.

Xanadu Research is dedicated to advancing 
various domains, with a strong emphasis on architectures~\citep{XanArch1,XanArch2,XanArch3,XanArch4,Borealis5,XanArch6,XanArch7,XanArch8,XanArch9,XanArch10,XanArch11,XanArch12,XanArch13,XanArch14,XanArch15,XanArch16,XanArch17,XanArch18,XanArch19,XanArch20,XanArch21}, 
hardware innovation~\citep{Borealis,XanHardw2,XanHardw3,XanHardw4,XanHardw5,XanHardw6,XanHardw7,XanHardw8,XanHardw9}, quantum programming~\citep{XanSoftw1,XanSoftw2,XanSoftw3,XanSoftw4,XanSoftw5,XanSoftw6,XanSoftw7,XanSoftw8,
XanSoftw9,Univ12-mod28,XanSoftw11,XanSoftw12,XanSoftw13,XanSoftw14,XanSoftw15,XanSoftw16,XanSoftw17,XanSoftw18,
XanSoftw19,XanSoftw20,XanSoftw21,XanSoftw22,XanSoftw23,XanSoftw24,XanSoftw25,XanSoftw26,XanSoftw27,XanSoftw28,XanSoftw29}, 
QML~\citep{XanQML1,XanQML2,XanQML3,XanQML4,XanQML5,XanQML6,XanQML7,XanQML8,XanQML9,XanQML10,XanQML11,XanQML12,XanQML13,XanQML14,XanQML15,XanQML16,XanQML17,XanQML18,XanQML19,XanQML20,XanQML21,XanQML22,XanQML23,XanQML24,XanQML25,XanQML26,XanQML27}
, and quantum chemistry~\citep{XanChem1,XanChem2,XanChem3,XanChem4,XanChem5,XanChem6,XanChem7,XanChem8,XanChem9,XanChem10,XanChem11,XanChem12}.  
The successful realization of quantum computational advantage using the dynamic programmability of \textit{Borealis} highlights the potential of photonic quantum computing and contributes substantially to the advancement of quantum computing technology.

\section{Quantum Photonics Industry}\label{Industry}

As reported by MarketsandMarkets™~\citep{PhotonicsMarket}, the global photonics market is anticipated to reach a valuation of USD 837.8 billion by 2025, reflecting a compound annual growth rate (CAGR) of 7.1\% from 2020 to 2025. This robust market growth is primarily fueled by the increasing adoption of photonics technologies across key sectors, including 
communication and information technologies, healthcare, and industrial manufacturing. The continued advancement and integration of photonics-enabled solutions within these industries are expected to drive sustained market expansion in the foreseeable future~\citep{PhotonicsMarket}.

The field of quantum communication is predominantly shaped by research and development efforts from industry and academic institutions. 
Analysis of patent applications between 2018 and 2021 highlights the leading entities actively involved in quantum communication~\citep{QcomReport2024}. Table~\ref{tab:qcom-patents} enumerates the top 20 institutions with the highest number of quantum communication-related patent applications during this period. Notable participants include technology giants such as Intel, Arqit, Huawei, LG Electronics, Toshiba, and QuantumCTek, alongside major telecommunications providers like Deutsche Telekom and British Telecom. Academic and research institutions such as Massachusetts Institute of Technology (MIT), Fraunhofer, Delft University of Technology, and South China Normal University~\citep{QcomReport2024}.

\begin{table}
\caption{Organizations leading in Quantum Communication-related, international patent applications during 2018-2021 (top 20)~\citep{QcomReport2024}.}
    \centering
    \resizebox{0.49\textwidth}{!}{
    \begin{tabular}{cccc}
        \hline \hline
        \textbf{No.} & \textbf{Institution} & \textbf{(2018-21)}\footnotemark[1] & \textbf{Country} \\
        \hline \hline
        1 & Intel  & 24 & United States \\
        \hline
        2 & Deutsche Telekom  & 23 & Germany \\
        \hline
        3 & Arqit  & 20 & United Kingdom \\
        \hline
        4 & Huawei Technologies Düsseldorf  & 20 & Germany \\
        \hline
        5 & LG Electronics  & 20 & South Korea \\
        \hline
        6 & Toshiba & 18 & Japan \\
        \hline
        7 & British Telecom  & 14 & United Kingdom \\
        \hline
        8 & QuantumCTek  & 14 & China \\
        \hline
        9 & Huawei Technologies  & 12 & China \\
        \hline
        10 & MIT  & 12 & United States \\
        \hline
        11 & Ericsson  & 12 & Sweden \\
        \hline
        12 & Fraunhofer & 11 & Germany \\
        \hline
        13 & IBM  & 11 & United States \\
        \hline
        14 & PsiQuantum  & 11 & United States \\
        \hline
        15 & Eagle Technology  & 10 & United States \\
        \hline
        16 & Delft University of Technology  & 10 & Netherlands \\
        \hline
        17 & Corning & 9 & United States \\
        \hline
        18 & ID Quantique  & 9 & Switzerland \\
        \hline
        19 & Microsoft  & 9 & United States \\
        \hline
        20 & South China Normal University & 9 & China \\
        \hline \hline
    \end{tabular}
    \label{tab:qcom-patents}
    \footnotetext[1]{Number of patent applications (2018-21)}
    }
\end{table}

\begin{table*}
\caption{\label{t:hardware} Additional photonic-centric quantum computing entities, sorted in ascending order based on year of foundation.}
\centering 
\begin{tabular}{c |p{1.2in} |p{3.7in} |c|c |l}
\hline \hline
No.&Company &Description/Technology   &Country &Founded      & Source\\
\hline
1  &Hamamatsu           &Hamamatsu engages in the production of a wide spectrum of devices encompassing silicon photodiodes, electron multipliers (EMs) designed for the detection of electrons, ions, and charged particles, along with multi-pixel photon counters. Moreover, Hamamatsu's photonics technology spans into the quantum information science (QIS) sector, offering a range of products tailored for applications such as quantum imaging, laser cooling/Bose-Einstein condensation (BEC), trapped ions, neutral atoms, and nitrogen vacancy cameras.
&Japan&1953&~\citep{Hamamatsu} \\
\hline
2  &Nordic Quantum Computing Group (NQCG)  &NQCG is building programmable optical quantum computers, aiming to attain quantum computational  advantage in simulation, optimization, and artificial intelligence. &Norway&2000 &~\citep{NQCG}\\
\hline 
3  &M Squared Lasers                       &Engineering pragmatic quantum devices, M Squared has undertaken the development of meticulously engineered laser systems, photonic components, and electrical instrumentation.  &UK& 2006&~\citep{MSquared}\\
\hline
4 &NTT Data                 &A multinational telecommunications corporation, owned by Nippon Telegraph \& Telephone (NTT) Corp., operates the Nanophotonics Center (NPC) founded in 2012. The NPC consists of research groups in NTT's laboratories, specializing in nanophotonics, including Nanodevices, Nanomechanics, Quantum Optical Physics, Photonic Nano-Structure, and Quantum Optical State Control. &Japan&2012&~\citep{NTTData} \\
\hline
5  &Cambridge Quantum Computing (CQC)\footnotemark[1]         &CQC is a British company focusing on quantum software and operating systems. &UK&2014&~\citep{Quantinuum} \\
\hline
6 &Sparrow Quantum                          &Sparrow Quantum design, develop, and produce high-quality Solid-state deterministic single-photon sources for quantum photonic applications and technologies. The Sparrow Quantum single-photon source relies on self-assembled InAs quantum dots coupled to a slow-light photonic-crystal waveguide, crafted using a distinctive processing technology~\citep{Sparrow1,Sparrow2,Sparrow3}. &Denmark&2015&~\citep{Sparrow} \\
\hline
7  &Ligentec                  &Ligentec specializes in the production of photonic integrated circuits (PIC) catering to a diverse array of industries and application areas, including:  
artificial intelligence (AI), 
quantum technologies
~\citep{Ligentec1,Ligentec2,Ligentec3,Ligentec4,Ligentec5,Ligentec6,Ligentec7,Ligentec8,Ligentec9,Ligentec10,Ligentec11,Ligentec12,Ligentec13,Ligentec14,Ligentec15,Ligentec16,Ligentec17,Ligentec18}, LiDAR~\citep{LiDAR1,LiDAR2,LiDAR3,LiDAR4,LiDAR5,Ligentec4,LiDAR7,LiDAR8}, 
nonlinear optics
~\citep{LigenticLOPTICS1,LigenticLOPTICS2,XanHardw4,XanHardw5,LigenticLOPTICS5,LigenticLOPTICS6,XanHardw3,LigenticLOPTICS8,LigenticLOPTICS9,LigenticLOPTICS10,LigenticLOPTICS11,LigenticLOPTICS12,Ligentec8,LigenticLOPTICS14,LigenticLOPTICS15,LigenticLOPTICS16,LigenticLOPTICS17,LigenticLOPTICS18,LigenticLOPTICS19,LigenticLOPTICS20,LigenticLOPTICS21,LigenticLOPTICS22}, 
optical frequency comb
~\citep{LigentecComb1,LigentecComb2,LigentecComb3,LigentecComb4,LigentecComb5,LigentecComb6,LigentecComb7,LigentecComb8,LigentecComb9,LigentecComb10,LigentecComb11,LigentecComb12,LigentecComb13,LigentecComb14,LigentecComb15,LigentecComb16,LigentecComb17,LigentecComb18,LigentecComb19,LigentecComb20},
quantum circuits~\citep{XanHardw2,XanHardw5,Ligentec3,LigentecQCirc4,LigentecQCirc5,Ligentec3,LigentecQCirc5},  
(De-)Muxes~\citep{LigenticDeMuxes1,LigenticDeMuxes2,LigenticDeMuxes3,LigenticDeMuxes4}, 
hybrid optics ~\citep{LigenticOptics1,LigenticOptics2,LiDAR3,LigenticOptics4}, 
spectroscopy~\citep{LigentecComb3,LIGENTICSpectroscopy2,LigentecComb7}, 
biosensors and light modulation and components~\citep{LigentecLight1,LigentecLight2,LiDAR1,LiDAR3,LiDAR4,LiDAR7,LigentecLight7,LigentecLight8,LigentecLight9,LigentecLight10,LigentecLight11,LigentecLight12,LigentecLight13}.
 &Switzerland&2016&~\citep{Ligentec} \\
\hline
8 &Vexlum                        & The company vision is to bring VECSELs (vertical external-cavity surface-emitting laser), optoelectronics processes, and laser systems for quantum technology applications~\citep{Vexlum1,Vexlum2,Vexlum3,Vexlum4,Vexlum5,Vexlum6,Vexlum7,Vexlum8,Vexlum9}.  &Finland&2017&~\citep{Vexlum} \\
\hline
9  &Nu Quantum                          &The company is actively involved in advancing single-photon components with a focus on the next generation of photonic quantum technologies. Their efforts extend to the development of end-to-end quantum cryptography systems, the establishment of quantum networks, and enhancing communication security. &UK&2018 &~\citep{NuQuantum}\\
\hline
10 &Quantropi                        &A quantum communications company pioneering quantum key distribution over the Internet. Quantropi has introduced its own novel post-quantum multivariate cryptosystem that establishes key exchange and digital signature mechanisms~\citep{Quantropi1,Quantropi2,Quantropi3,Quantropi1,Quantropi5,Quantropi6,Quantropi7,Quantropi8,Quantropi9,Quantropi10,Quantropi11,Quantropi12,Quantropi13,Quantropi14,Quantropi15,Quantropi16,Quantropi17,Quantropi18,Quantropi19,Quantropi20}. &Canada&2018&~\citep{Quantropi} \\
\hline
11  &Aegiq                               &Aegiq specializes in quantum computing and networking, utilizing hybrid integrated photonics to create scalable quantum applications, from computing to network interconnects and cryptographic communications.  
Aegiq introduced deterministic photon sources, namely iSPS for photonic quantum computing~\citep{iSPS}.
&
 UK&2019&~\citep{Aegiq} \\
\hline
12  &Miraex                    &Miraex develops photonic sensing and distributed quantum computing solutions~\citep{Miraex}.  &Switzerland&2019&~\citep{Miraex}\\
\hline
13  &QBoson                    &QBoson dedicated to the development of scalable and programmable optical quantum computing platforms, as well as the implementation of practical quantum computing applications. The company has successfully constructed a cutting-edge optical quantum computer, named ``Tiangong Quantum Brain," boasting a qubit capacity of 100~\citep{Enlighten44,Enlighten}.  &China&2020&~\citep{QBoson}\\
\hline \hline
\end{tabular}
\footnotetext[1]{CQC engaged in a merger with Quantinuum on November 30, 2021.}
\end{table*}

The utilization of photonic technology is increasingly propelling the global photonics market's expansion, primarily owing to its exceptional versatility and utility in various industrial applications.  
Industries, such as aerospace and defense, are increasingly embracing silicon photonics due to the compact form factor, reduced power consumption, and lighter weight of photonics-based devices, which additionally exhibit robust immunity to electromagnetic interference, ensuring consistent performance. With the burgeoning development of autonomous vehicle technology, photonics technology is poised to play a pivotal role in car-to-car communication, navigation, and mapping. 
The growing adoption of photonic technology across diverse sectors significantly contributes to the global photonics market's growth. Notably, this trend holds particular significance for the quantum photonic technology industry. 
Table~\ref{t:hardware} summarizes various innovators and developers in the field of photonic quantum technologies.

%% file: Acknowledgements.tex
\section*{Declaration of competing interest}

The author declare that they have no known competing financial interests or personal relationships that could have appeared to influence the work reported in this paper.

\subsection*{Ethical Approval and Consent to participate}
Not applicable.

\subsection*{Consent for publication}
The author has approved the publication. This research did not involve any human, animal or other participants.

\subsection*{Availability of data and materials}
The datasets generated during and/or analyzed during the current study are included within this article.

\subsection*{Competing interests}
The author declare no competing interests.

\subsection*{Funding}

The authors declare that no funding, grants, or other forms of support were received at any point throughout this research work. 
“This research received no specific grant from any funding agency in the public, commercial, or not-for-profit sectors.”

\subsection*{Acknowledgements}

I am deeply grateful to the crowns of honor, though words cannot fully express my appreciation. 
The views and conclusions expressed in this work are solely those of the author and do not necessarily reflect the official policy or position of:  iPronics Programmable Photonics, The USTC Jiuzhang light-based quantum computers, ORCA Computing, PsiQuantum,  
Quandela Photonic Quantum Computers, The QuiX Quantum, TundraSystems Global, TuringQ, Photonic Inc, Xanadu Quantum Technologies, 
Intel, Deutsche Telekom, Arqit, Huawei Technologies Düsseldorf, LG Electronics, Toshiba, British Telecom, QuantumCTek, Huawei Technologies, MIT, Ericsson, Fraunhofer, IBM, Rigetti, Eagle Technology, Delft University of Technology, Corning, ID Quantique, Microsoft, South China Normal University, Hamamatsu, Nordic Quantum Computing Group (NQCG), M Squared Lasers, NTT Data, Cambridge Quantum Computing (CQC), Sparrow Quantum, Ligentec, Vexlum, Nu Quantum, Quantropi, Aegiq, Miraex, or any affiliated organizations.

\subsection*{Data availability}

The data sets produced and/or analyzed during the current study are incorporated within this article.